\documentclass[12pt,english]{article}

\usepackage{mathptmx}
\usepackage{geometry}
\usepackage{xcolor}
\usepackage{array}
\usepackage{bbding}
\usepackage{multirow}
\usepackage[fleqn]{amsmath}
\usepackage{amssymb}
\usepackage{amsthm}
\usepackage{graphicx}
\usepackage{natbib}
\usepackage{lscape}
\usepackage[hyphens]{url}
\usepackage[hidelinks]{hyperref}
\usepackage{comment}
\usepackage{bm}
\usepackage[title]{appendix}
\usepackage{longtable}
\usepackage{mathtools}
\usepackage{chngcntr}
\usepackage{authblk}
\usepackage{babel}
\usepackage{dsfont}
\usepackage{subcaption}
\usepackage{fancyhdr}
\usepackage{cleveref}
\usepackage{abstract}
\usepackage{float}
\usepackage{afterpage}
\usepackage{csquotes}
\usepackage{placeins}
\usepackage{enumitem}

\usepackage[ruled,vlined]{algorithm2e} 

\makeatletter
\allowdisplaybreaks
\date{}

\fancypagestyle{plain}{%
	\fancyhf{} 
	
	\lhead{\footnotesize \textcolor{gray}{}\ \ \textcolor{gray}{\textbf{}}}
	\rhead{\footnotesize \textcolor{gray}{\thepage}}
}
\makeatletter
\def\blfootnote{\xdef\@thefnmark{}\@footnotetext}
\makeatother

\makeatletter
\def\titlepageext{
		
    {\small {\bf Abstract: }\abstract \\

    \small {\bf Keywords:} \keyword
    }

	\blfootnote{* Corresponding author. E-mail address: \href{mailto:\corresemail}{\corresemail}.}
}
\makeatother

\usepackage[mathlines, pagewise]{lineno}
\usepackage{etoolbox} 

\newcommand*\linenomathpatchAMS[1]{%
	\expandafter\pretocmd\csname #1\endcsname {\linenomathAMS}{}{}%
	\expandafter\pretocmd\csname #1*\endcsname{\linenomathAMS}{}{}%
	\expandafter\apptocmd\csname end#1\endcsname {\endlinenomath}{}{}%
	\expandafter\apptocmd\csname end#1*\endcsname{\endlinenomath}{}{}%
}

\expandafter\ifx\linenomath\linenomathWithnumbers
\let\linenomathAMS\linenomathWithnumbers
\patchcmd\linenomathAMS{\advance\postdisplaypenalty\linenopenalty}{}{}{}
\else
\let\linenomathAMS\linenomathNonumbers
\fi

\linenomathpatchAMS{gather}
\linenomathpatchAMS{multline}
\linenomathpatchAMS{align}
\linenomathpatchAMS{alignat}
\linenomathpatchAMS{flalign}

\nolinenumbers

\title{Joint Estimation of Dynamic O-D Demand and Choice Models for Dynamic Multi-modal Networks: Computational Graph-Based Learning and Hypothesis Tests}

\author[a]{Xiaoyu Ma}
\author[a,b,$\ast$]{Sean Qian}
\affil[a]{Department of Civil and Environmental Engineering, Carnegie Mellon University, Pittsburgh, PA 15213, US}
\affil[b]{Heinz College of Information Systems and Public Policy, Carnegie Mellon University, Pittsburgh, PA 15213, US}

\def\corresemail{seanqian@cmu.edu}

\def\abstract{Understanding travel demand and behavior, particularly route and mode choices, is critical for effective transportation planning and policy design in multi-modal systems with emerging mobility options. Multi-modal system-level data, such as traffic counts, probe speeds, and transit ridership, offer scalable, cost-effective, and privacy-preserving advantages for inferring and analyzing travel behavior. This research use such system-level data to infer travel demand and choices that vary by time of day, origin/destination location and mode. Existing studies focus on a single transportation mode, consider limited behavioral factors in disutility functions, rely on static travel time functions, and face computational challenges when applied to large-scale networks. This research addresses these gaps by proposing a joint estimation framework for dynamic origin–destination (O–D) demand and disutility functions within a multi-modal transportation system that includes both private driving and public transit, using multi-source system-level data. A multi-modal dynamic traffic assignment model that accounts for both route and mode choices is integrated into the framework, with detailed travel time modeling for multiple modes. Alternative-specific and zone-specific factors are incorporated into generic disutility functions to capture heterogeneous traveler perceptions. The estimation problem is formulated and solved using a computational graph–based approach, enabling dynamic network modeling and scalable inference across large-scale networks and diverse data sets. Furthermore, we propose a hypothesis testing framework tailored to this complex estimation setting to assess the statistical significance of behavioral parameters, thereby enabling model selection and statistically rigorous insights for real-world applications.}

\def\keyword{Dynamic O-D demand; Disutility function; Multi-modal systems; Computational graph; Hypothesis tests; Multi-source system-level data; Boarding and alighting count }

\begin{document}
\maketitle
\titlepageext

\section{Introduction}
Learning how travelers make different route and mode choices is critical, particularly in the multi-modal transportation context with emerging mobility options. The estimation results and derived insights provide crucial policy implications that can guide the planning process, policy design, and transportation system operations. Over the decades, many widely adopted approaches have been based on data from individual travel surveys and location-based service (LBS) data. Such approaches can yield satisfactory estimates and provide valuable insights. However, survey-based approaches suffer from several drawbacks: stated preferences may not necessarily align with real-life choices, potential sampling biases, high data collection costs, and limited coverage in both spatial and temporal dimensions. LBS data are passively collected and require relatively low costs, but they may contain behavioral biases and require intensive pre-processing for trip extraction \citep{zheng2015trajectory, wang2019extracting, wang2018data, pan2024national, lu2024two, wang2025exploring}.

System-level data, such as probe-vehicle speeds, passenger counts, traffic counts, and densities, advanced by the rapid development of sensing technology, cloud computing, and smart devices, may overcome the disadvantages of individual-level survey or LBS data and are becoming increasingly valuable for understanding network flow, including characteristics of origin–destination (O-D) trips and choice behavior. The main advantages are low cost and ubiquitous spatio-temporal coverage while preserving privacy. System-level data, which capture aggregated features of traffic flow such as time-varying traffic counts and probe speeds, have become widely available at large spatial-temporal scales thanks to advanced technologies. Examples of probe-speed providers include INRIX, HERE, TomTom, and Google Maps. Furthermore, multiple sources of data are now widely available for complex multi-modal transportation systems. For example, traffic count/speed data (by vehicle classification) from sensors, traffic density data from images and videos, transit ridership data from smart cards, location/trajectory data from smart devices on private/public vehicles, and micromobility usage data at the station level. How to leverage multi-source system-level data simultaneously for effective and accurate estimation of network flow and travel behaviors across multiple modes remains unclear and is the focus of this research paper.

\subsection{Literature review and research gaps}

In the literature, a rich body of studies has utilized system-level data for various estimation tasks in single-mode systems, including O-D demand estimation for private driving, estimation of travelers' utility or disutility functions, joint estimation of car O-D demand and behavioral parameters, and calibration of link performance functions. In addition, the estimation of public transit O-D demand is often studied separately from the estimation of private driving O-D demand, without considering the interactions among different transportation modes or the shared impacts resulting from traffic congestion.

Because the existing literature mainly focuses on single-mode systems, the following provides a brief summary of the relevant studies by category and highlights the corresponding research gaps.

A branch of studies focuses on the estimation of private driving O-D demand, for example, \cite{Fisk1989, Yang1992, Cascetta2001, Wu2018, Ma2018, Krishnakumari2020, Ma2020, Wollenstein2022, cao2024dynamic, vahidi2025time}. Other studies examine the estimation of travelers' utility or disutility functions \citep{Robillard1974, Fisk1977, Daganzo1977, Anas1990, Liu1996, Arriagada2022, Dixit2024, GuardaandQian2024}. There are also studies focusing on the joint estimation of private driving O-D demand and behavioral parameters \citep{Yang2001, Lo2003, García-Ródenas2009, Russo2011, Wang2016, Guarda2024}. Finally, some studies investigate the calibration of static link performance functions \citep{Russo2011, García-Ródenas2013, Suh1990, Wollenstein2022}.

In the public transit domain, the estimation of transit O-D demand is typically conducted separately from the estimation of private driving O-D demand and often focuses on a static network or a limited set of selected transit lines. These studies generally rely on Automatic Passenger Counting (APC) or Automatic Vehicle Location (AVL) data to estimate transit O-D demand \citep{li2007generalized, ji2015transit, sun2021estimating, galliani2024estimation, tang2024origin, saeidi2024passenger, zhao2024origin, CHEN2025103278}. The methodologies used in this literature include statistical models such as Bayesian inference \citep{sun2021estimating, CHEN2025103278}, gravity model calibration \citep{saeidi2024passenger, zhao2024origin}, Iterative Proportional Fitting \citep{galliani2024estimation}, and machine learning approaches \citep{tang2024origin}.
  
The literature has significantly advanced the estimation and analysis of travel demand and behavior from various perspectives within single-mode transportation systems. However, there remain research gaps that need to be addressed or can be further improved in the context of multi-modal transportation systems, as summarized below. 
\begin{enumerate}
  \item Existing studies largely focus on single-mode cases, with limited attention to multi-modal systems that incorporate various modes such as private cars, transit, park-and-ride, and others. This gap may stem from the inherent complexity of modeling multi-modal systems and the historical lack of integrated multi-modal data. In reality, however, different modes are interdependent. Namely, cars and buses share the road infrastructure and affect each other’s travel time, and travelers’ mode choices usually have inter-modal correlations. With the increasing availability of multi-source, high-resolution system-level data, it is both feasible and necessary to consider multiple modes jointly to understand how travelers simultaneously make mode and route decisions that vary by time of day, location, and population characteristics.
  
  \item Most studies that estimate travelers' (dis)utility functions using system-level data (instead of individual surveys) consider an highly limited set of factors, e.g., travel time only, meanwhile without alternative-specific parameters \citep[e.g.,][]{Robillard1974, Fisk1977, Daganzo1977, Anas1990, Liu1996, Yang2001, Lo2003, García-Ródenas2009, Russo2011, Wang2016, Wu2018, Ma2020}. However, travelers tend to perceive differently for different alternatives even with respect to the same factor; therefore, adopting alternative-specific parameters may further advance understanding of the travel behavior in the network context. For example, travelers using different modes may have varying sensitivity with respect to the change in travel time categorized in the disutility associated with their respective modes. Additionally, multi-source data provides opportunities to specify a more sophisticated (dis)utility function involving multiple factors, both alternative-specific and zone-specific, beyond travel time (such as monetary and sociodemographic factors), enabling more comprehensive and in-depth analysis.
  \item Much of the existing research adopts simplified assumptions that travel time is invariant across time of day, thereby overlooking the dynamic nature of congestion (e.g., queuing, spillover effects) or relies on static link performance functions that may not reflect real-world traffic conditions. In practice, travel times vary significantly due to within-day traffic dynamics, passenger loading and unloading patterns on buses, and other temporal factors. These variations should be explicitly incorporated into a dynamic network context to produce realistic estimates of traveler behavior and system performance.
  \item There is a lack of hypothesis-testing frameworks for complex, non-closed-form estimation problems in dynamic multi-modal systems. After estimating behavioral parameters, it is essential to assess whether the estimated parameters or associated factors are statistically significant, as this supports feature selection, model selection, and behavioral interpretation. These processes are fundamental for achieving robust and generalizable behavior models that can be applied to large-scale multi-source datasets. Among existing studies, only \citep{GuardaandQian2024} performs hypothesis testing; however, their analysis is restricted to single-mode cases and relies on a static link performance function with a limited set of factors. The present paper extends hypothesis testing to a more general setting involving multi-modal and dynamic networks with richer disutility specification. 
  \item Computational burden remains a major challenge in the literature and often limits the size of networks and datasets that can be analyzed. Given the complexity of the problem addressed in this paper, we develop a generic computational-graph-based approach to overcome these computational challenges and enable efficient estimation for large-scale networks.
\end{enumerate}

\subsection{Contributions}
This paper intends to develop a holistic framework for the joint estimation of multi-modal dynamic O–D demand and choice models (i.e., parameters in the disutility functions), along with hypothesis testing on the estimated parameters. The multi-modal system includes private car, public transit (bus and metro) services, and the combined private and public transport modes. The proposed framework integrates a mesoscopic multi-modal dynamic traffic simulation and a computational graph-based learning approach, minimizing the discrepancies between simulated outcomes and the observations (i.e., dynamic link flow, link travel time, and boarding/alighting counts at the trip and stop levels). 

Addressing the research gaps and challenges presented above, the main contributions of this research include the following.
\begin{enumerate} 
  
  \item \textbf{Multi-modal consideration and a holistic mesoscopic simulation-based approach.} This research proposes a framework for the joint estimation of dynamic O–D demand and travel disutility functions for a dynamic multi-modal transportation system involving multiple modes, including car, bus, metro, and park-and-ride. Building on a multi-modal dynamic traffic assignment model that incorporates both mode choice and route choice, this research leverages system-level data including traffic counts, transit ridership, and travel time to jointly estimate dynamic O–D demand and the parameters of travelers' disutility functions. Additionally, while most existing research infers public transit O–D demand from trip-based boarding/alighting data and focuses exclusively on a single mode, a static network, or selected transit lines, this research offers a holistic network(or city)-level approach that leverages multi-modal, multi-source data. 
  \item \textbf{Hierarchical mode/route choice model.} This research tailors a hierarchical mode choice structure considering the distinct features of different travel modes that include a wide range of travel options and combinations of multiple options used across different lags of a trip. This enables more meaningful, generic, and comprehensive mode choice behavioral modeling in a more granular manner.
  \item \textbf{Generic disutility function with both alternative-specific and zone-specific factors.} This research leverages multi-source data to incorporate a wide range of factors that may influence travelers' behaviors, including both alternative-specific and zone-specific factors. In addition to travel time, factors such as waiting time, monetary cost, comfort level, and sociodemographic characteristics (e.g., income level, population density) are considered in the disutility function. Moreover, alternative-specific parameters are employed for selected factors based on their physical interpretation. 
  \item \textbf{Computational graph-based learning approach.} This research casts the joint dynamic O-D demand and disutility function estimation problem on a computational graph to solve it. Leveraging computational graph-based learning, this approach addresses several challenges: (a) the ability to handle large-scale networks and high-dimensional variables/parameters; (b) the capability to model dynamic travel times (i.e., dynamic network loading) to capture congestion effects; and (c) the flexibility to incorporate and utilize any combination of multi-source data (even with missing data) for network calibration.
  \item \textbf{Hypothesis testing framework for complex non-closed-form estimation problems.} A framework for statistical hypothesis tests on the estimated parameters is proposed, allowing insights into the significance of different factors and relevant policy implications for real-world tasks. To address challenges posed by the non-closed-form nature of traffic dynamics and the high dimensionality of the estimation problem, we employ an approach based on the Wald principle while leveraging the computational graph.
\end{enumerate}

The remainder of this paper is organized as follows. Section \ref{sec:model} presents the models, including the multi-modal network building, the multi-source data-driven mode choice model, and the multi-modal dynamic traffic assignment. Section \ref{sec: dode_cg} proposes the problem formulation of the joint estimation of dynamic O-D demand and disutility parameters, and the computational graph-based approach for solving it. Section \ref{sec: hypo} presents the hypothesis testing framework designed for our problem. Section \ref{sec: experiments} presents the experiments on two different networks. Section \ref{sec-Discussions} provides insightful discussions. Section \ref{sec: conclusions} concludes the paper. 

\section{Model} \label{sec:model}

\subsection{Multi-modal transportation network} \label{sec:multimodal network}
\begin{figure}[ht]
    \centering
    \includegraphics[width=\linewidth]{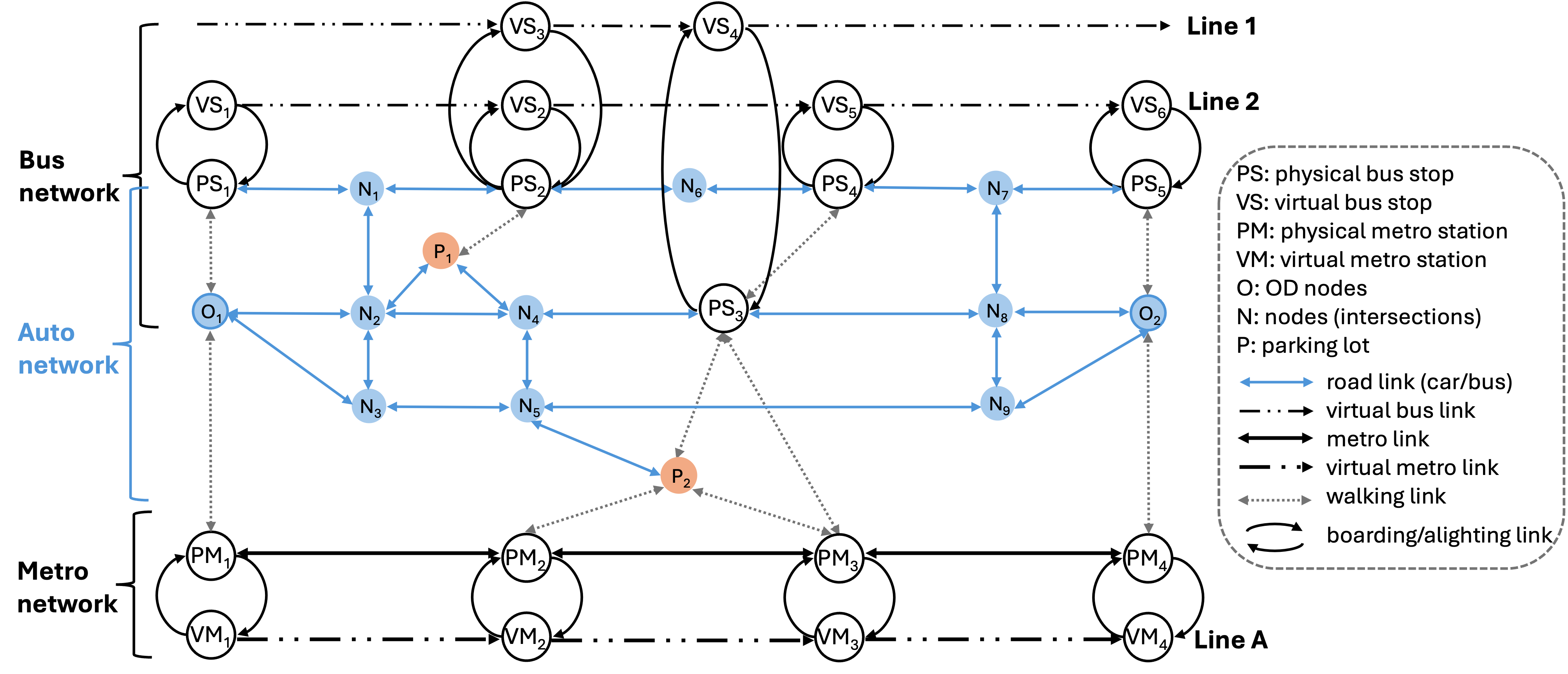}
    \caption{A combined multi-modal transportation network}
    \label{fig-combined network}
\end{figure}

In this research, we assume that travelers can drive a car, take a bus, take the metro, or use any combination of these modes including park-and-ride (PNR). To accommodate multiple modes, a multi-modal transportation network that captures the characteristics of each mode as well as the interconnections among them needs to be established. Following the combined network structure proposed in the previous work \citep{Pi2019}, \Cref{fig-combined network} shows a schematic network for the multi-modal transportation system in this research.

The auto network represents the network used by cars. Buses use portions of the auto network depending on their route sets. The bus network also includes stops, consisting of physical stops (PS) and virtual stops (VS). A single physical bus stop may correspond to multiple virtual stops, each associated with a bus line that serves that location. Links between virtual stops represent bus movements between stops, while links between a virtual stop and its corresponding physical stop represent travelers’ boarding and alighting behaviors.

Similarly, the metro network includes physical metro stations (PM) and virtual metro stations (VM) to model metro line operations and travelers’ boarding and alighting behaviors. The key difference is that the metro network is exclusively used by metro services, and the travel times between metro stations are fixed because metro trips are not affected by road traffic congestion.

Parking facilities are represented as parking nodes in the auto network. Walking links connect physical bus stops (or metro stations) to origin/destination nodes, physical bus stops (or metro stations) to parking nodes, and different physical bus stops or metro stations to each other. These walking links connect the sub-networks for car, bus, and metro, enabling transfer behaviors and combination uses of the modes.

\subsection{Multi-source data-driven mode choice model for network flow}
\subsubsection{Nested structure design}
\begin{figure}[ht]
    \centering
    \includegraphics[width=0.5\textwidth]{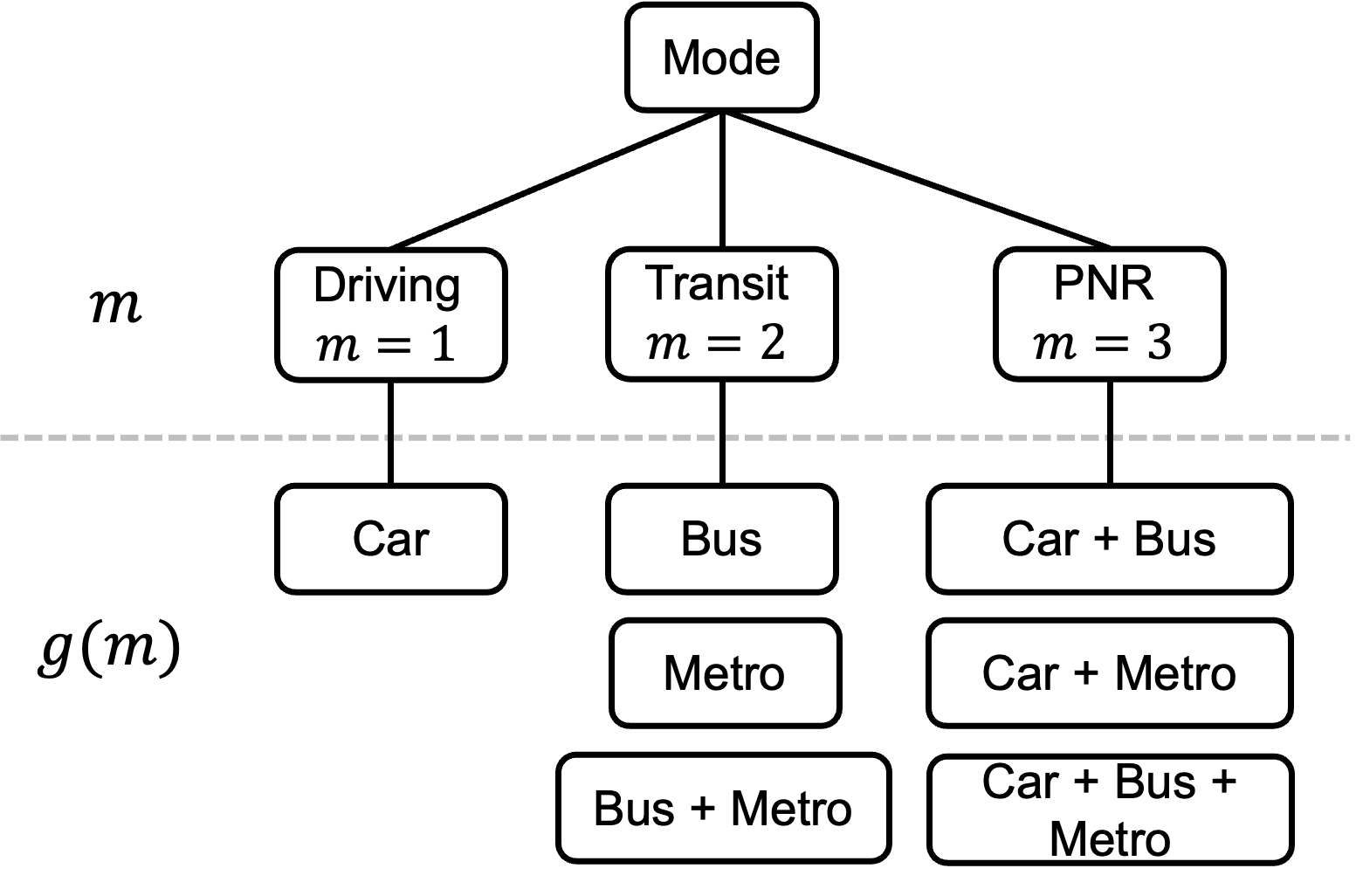}
    \caption{Nested mode choice structure}
    \label{fig-nested structures}  
\end{figure}

In this research, a nested structure with two layers is proposed to model travelers’ mode choice behaviors. In the first layer, travelers choose among three general options: driving, public transit, and PNR. Within the public transit nest, the second layer includes three options: bus only, metro only, and a combination of bus and metro. Correspondingly, within the PNR nest, the second-level options are car+bus, car+metro, and car+bus+metro. The nested structure is illustrated in \Cref{fig-nested structures}.

Note that not all modes need to be available for every O–D pair. The proposed formulations and models in this paper apply to the general case in which some modes may be unavailable for certain O–D pairs.

\subsubsection{Generalized disutility functions with alternative-specific and zone-specific variables}
Leveraging multi-source data, the proposed disutility functions include a more comprehensive set of variables, encompassing both alternative-specific and zone-specific variables, designed for the mesoscopic simulation-based modeling approach. The alternative-specific variables include travel time, waiting time, parking fees, and transit fares, which capture the effects of different alternatives due to their distinct characteristics. The zone-specific variables reflect sociodemographic factors and account for inherent heterogeneity among travelers with different backgrounds. For example, travelers with different income levels may have different mode preferences, and individuals living in highly populated areas may prefer public transit over driving. In this research, we use median income and population density at the Traffic Analysis Zone (TAZ) level as the two sociodemographic factors. Additional factors can be incorporated by extending the disutility functions.

Another important consideration is the use of alternative-specific parameters for the same variable across different modes when such distinctions are physically meaningful. For example, (1) changes in travel time may produce different levels of disutility for car drivers compared to transit users; and (2) changes in waiting time may have different impacts on bus versus metro travelers, since the conditions of waiting environments differ (e.g., metro stations are less affected by weather).

As shown in \Cref{fig-nested structures}, let \(m \in M=\{1,2,3\}\) denote the first-level mode choice and \(g(m) \in G(m)\) denote the second-level mode choice within nest \(m\), where \(G(m)\) is the set of all second-level choices in nest \(m\) and \(M\) is the set of all first-level mode choices. Denote O-D pair \((r,s) \in \Omega\) where \(\Omega\) represents the set of O-D pairs. Denote \(c^{rs}_{m,g(m),k,t}\) as the generalized travel cost/disutility from origin \(r\) to destination \(s\) taking first-level mode \(m\) and second-level mode \(g(m) \in G(m)\) using path \(k \in P^{rs}_{m,g(m)}\) departing at time \(t \in T\), where \(P^{rs}_{m,g(m)}\) is the path set for mode \(g(m)\) from \(r\) to \(s\) and \(G(m)\) is the set of all the second-level mode choices under choice \(m\).  Denote \(\gamma_{1,m,g(m)}\), \(\gamma_{2,m,g(m)}\), and \(\gamma_{3,m,g(m)}\) as the associated coefficient of income, population density of origin, and population density of destination, respectively, for mode \(g(m) \in G(m), m \in M\). And denote \(\alpha_{m,g(m)}, g(m) \in G(m), m \in \{2,3\}\) as the inherent inconvenience cost compared to the mode of driving, and the higher the value, the higher the inconvenience.

Then the disutility of driving using path \(k\) and departing at time \(t\) is
\begin{equation}
\begin{aligned}
    c^{rs}_{1,car,k,t} = \beta_{1,car} w^{rs}_{k,t,car} + \beta_2 \tau_k^{rs} + \gamma_{1,1,car} I^{rs} + \gamma_{2,1,car} J^{r} + \gamma_{3,1,car} J^{s}, \forall k \in P^{rs}_{1,car}, t \in T
\end{aligned}
\end{equation}
where
\begin{itemize}
    \item \(w^{rs}_{k,t,car}\) is the driving travel time using path \(k\) departing at time \(t\). The associated parameter \(\beta_{1,car}\) is alternative-specific.
    \item \(\tau_k^{rs}\) is the average parking fee if taking path \(k\).
    \item \(I^{rs}\) is the median income of travelers from \(r\) to \(s\). The associated parameter \(\gamma_{1,1,car}\) is alternative-specific.
    \item \(J^{r}\) and \(J^{s}\) represent the population density of the origin \(r\) and destination \(s\), respectively. The associated parameters \(\gamma_{2,1,car}\) and \(\gamma_{3,1,car}\) are alternative-specific. 
    \item There is no constant term in the disutility function for driving because, as defined earlier, the driving mode is treated as the baseline. The constant \(\alpha_{m,g(m)}, g(m) \in G(m), m \in \{2,3\}\) represents the additional inconvenience cost of choosing another mode compared to the driving mode.
\end{itemize}

The disutility of taking bus only, metro only, or bus + metro,  (i.e., \(c^{rs}_{2,bus,k,t}\),\(c^{rs}_{2,metro,k,t}\),\(c^{rs}_{2,bus+metro,k,t}\)), using path \(k\) and departing at time \(t\) is formulated as
\begin{equation}
\begin{aligned}
    c^{rs}_{2,g(2),k,t} = & \xi_{bus}^{g(2)}(\beta_{1,bus} w^{rs}_{k,t,bus} +\beta_{3,bus} \bar{w}^{rs}_{k,t,bus}) + \xi_{metro}^{g(2)}(\beta_{1,metro} w^{rs}_{k,t,metro}  + \beta_{3,metro}\bar{w}^{rs}_{k,t,metro}) \\
    & + \beta_2 \delta_k^{rs} + \beta_4 \tilde{w}^{rs}_{k,t} +\gamma_{1,2,g(2)} I^{rs} + \gamma_{2,2,g(2)} J^{r} + \gamma_{3,2,g(2)} J^{s} + \alpha_{2,g(2)}, \\
    & \forall k \in P^{rs}_{2,g(2)},g(2) \in G(2), t \in T
\end{aligned}
\end{equation}
where 
\begin{itemize}
    \item \(\xi_{bus}^{g(m)}\) and \(\xi_{metro}^{g(m)}\), \(m \in \{2,3\}\), are binary indicators representing whether the mode \(g(m)\) involves bus or metro trip, respectively, with the value being 1 representing yes and 0 otherwise.
    \item \(w^{rs}_{k,t,bus}\) and \(w^{rs}_{k,t,metro}\) represent the travel time in the bus and metro train, respectively, when taking the path \(k\) and departing at time \(t\). The associated parameters \(\beta_{1,bus}\) and \(\beta_{1,metro}\) are alternative-specific.
    \item \(\bar{w}^{rs}_{k,t,bus}\) and \(\bar{w}^{rs}_{k,t,metro}\) represent the waiting time at bus stops and metro stations, respectively, when taking the path \(k\) and departing at time \(t\).  The associated parameters \(\beta_{3,bus}\) and \(\beta_{3,metro}\) are alternative-specific.
    \item \(\delta_k^{rs}\) is the total transit fare of transit route \(k\).
    \item \(\tilde{w}^{rs}_{k,t}\) is the walking time when taking transit route \(k\) and departing at time \(t\).
    \item \(\alpha_{2,g(2)}\) is the constant term representing the inconvenience as defined earlier.
    \item Other variables have the same meaning as previously presented.
\end{itemize}

The disutility of choosing either of the PNR options (i.e., \(c^{rs}_{3,car+bus,k,t}\),\(c^{rs}_{3,car+metro,k,t}\), and \(c^{rs}_{3,car+bus+metro,k,t}\)) using path \(k\) and departing at time \(t\) is formulated as
\begin{equation}
\begin{aligned}
    c^{rs}_{3,g(3),k,t} = & \beta_{1,car} w^{rs}_{k,t,car} + \beta_2 \tau_k^{rs} + \xi_{bus}^{g(3)}(\beta_{1,bus} w^{rs}_{k,t,bus} +\beta_{3,bus} \bar{w}^{rs}_{k,t,bus}) + \xi_{metro}^{g(3)}(\beta_{1,metro} w^{rs}_{k,t,metro} \\
    & + \beta_{3,metro}\bar{w}^{rs}_{k,t,metro}) + \beta_2 \delta_k^{rs} + \beta_4 \tilde{w}^{rs}_{k,t} + \gamma_{1,3,g(3)} I^{rs} + \gamma_{2,3,g(3)} J^{r} + \gamma_{3,3,g(3)} J^{s} + \alpha_{3,g(3)}, \\
    & \forall k \in P^{rs}_{3,g(3)},g(3) \in G(3), t \in T
\end{aligned}
\end{equation}

So far, the disutility functions for all \(g(m) \in G(m), m \in M\) have been presented. Denote vectors \(\pmb{\beta} = \{\beta_{1,car}, \beta_{1,bus}, \beta_{1,metro}, \beta_2, \beta_{3,bus}, \beta_{3,metro}, \beta_4\}\), \(\pmb{\gamma} = \{\gamma_{1,m,g(m)}, \gamma_{2,m,g(m)},\gamma_{3,m,g(m)}, g(m)\in G(m), m \in M\}\), and \(\pmb{\alpha} = \{\alpha_{m,g(m)}, g(m) \in G(m), m \in \{2,3\}\}\). The parameters (\(\pmb{\beta}\)) associated with alternative-specific variables, the parameters (\(\pmb{\gamma}\)) associated with zone-specific variables, and the alternative-specific constants \(\pmb{\alpha}\) are all the parameters in disutility functions that need to be estimated.

According to random utility theory and the nested logit model, the probability of choosing the first-level choice of nest \(m\) for a traveler in the O-D pair \((r,s)\) with departure time \(t\) can be calculated by the following \Cref{1st level prob}.
\begin{equation}
    \label{1st level prob}
    \mathcal{P}_{m,t}^{rs} = \frac{e^{\theta_m IV_{m,t}^{rs}}}{\sum_{m \in M}e^{\theta_m IV_{m,t}^{rs}}}
\end{equation}
where \(\theta_m\) is the scale parameter for the nest \(m\), and \(IV_{m,t}^{rs}\) is the inclusive value (log-sum term) of the nest \(m\) that can be calculated by \Cref{logsum}.
\begin{equation}
    \label{logsum}
    IV_{m,t}^{rs} = \ln(\sum_{g(m) \in G(m)} e^{-\frac{1}{\theta_m}c^{rs}_{m,g(m),t}})
\end{equation}
where \(c^{rs}_{m,g(m),t}\) is the overall cost of choosing mode \(g(m)\) for a traveler in the O-D pair \((r,s)\) with departure time \(t\).

Then within the nest \(m\), the probability of choosing mode \(g(m)\) for a traveler in the O-D pair \((r,s)\) with departure time \(t\) is calculated as,
\begin{equation}
    \mathcal{P}_{g(m)|m,t}^{rs} = \frac{e^{-\frac{1}{\theta_m}c^{rs}_{m,g(m),t}}}{\sum_{j \in G(m)} e^{-\frac{1}{\theta_m}c^{rs}_{m,j,t}}}.
\end{equation}

Then the overall probability of choosing mode \(g(m)\) for a traveler in the O-D pair \((r,s)\) with departure time \(t\) is as follows.
\begin{equation}
     \mathcal{P}_{g(m),t}^{rs} = \mathcal{P}_{m,t}^{rs} \cdot \mathcal{P}_{g(m)|m,t}^{rs}
\end{equation}

If certain modes are unavailable for a given O–D pair, their corresponding choice probabilities will be zero, and these modes will be excluded from the choice set.

It is important to separate different types of time-related costs and to use alternative-specific variables in the disutility function. First, travelers perceive different levels of disutility for different types of time (such as in-vehicle travel time, waiting time, and walking time) and these perceptions vary with factors like in-vehicle comfort, waiting-area environment, and weather conditions. Separating these components enables more granular modeling and analysis of travel behaviors and their influencing factors. Second, even for the same type of cost, it is important to use alternative-specific parameters because disutility may be perceived differently across modes. For example, increases in travel time or waiting time may lead to smaller increases in disutility for metro users than for bus users, since metro services typically offer better in-vehicle and waiting conditions. Using alternative-specific parameters isolates the distinct impacts of a factor across modes, facilitating analysis of intermodal relationships and relevant policy implications.

The disutility functions presented above are illustrative examples that include different types of time costs (travel time, waiting time, walking time), monetary costs (parking fees, transit fares), and sociodemographic factors (income, population density). Other factors can be incorporated into the generic formulation, and the proposed framework for joint estimation of dynamic O–D demand and choice models is applicable to a wide range of potential variables.

\subsection{Multi-modal dynamic equilibrium: a generalized three-layer nested mode-route choice model}

Within each mode, travelers make route choices. Given the proposed two-layer nested logit model for mode choice, we assume that the route choice behavior within each mode similarly follows the logit model based on random utility theory. The routing choice modeling here is essentially the logit-based stochastic dynamic user equilibrium. The behavioral foundation of stochastic user equilibrium is that each traveler attempts to minimize their perceived travel cost/disutility which is composed of a deterministic measured cost/disutility and a random term that can be interpreted as perceptual error \citep{daganzo1977stochastic, hazelton1998some}. Different rules, such as logit-based or probit-based formulations, have been applied in modeling stochastic dynamic user equilibrium \citep{han2003dynamic, lim2005dynamic, zhang2013dynamic, long2015nonlinear}. In this research, we adopt the logit-based rule and introduce a third layer to the nested structures shown in \Cref{fig-nested structures}, representing logit-based route choices under each second-layer mode choice. Consequently, the multi-modal dynamic equilibrium is modeled using a three-layer nested mode–route choice model.

Denote \(h^{rs}_{m,g(m),t}\) as the flow of second-level mode \(g(m)\) under \(m\) in O-D pair \((r,s)\) with departure time \(t\). And \(q^{rs}_{t}\) is the total demand in O-D pair \((r,s)\) with departure time \(t\). Let \(f^{rs}_{m,g(m),k,t}\) represent the path flow of path \(k\) with mode choice \(g(m)\) in O-D pair \((r,s)\) with departure time \(t\). The proposed multi-modal dynamic equilibrium can be formulated as the following set of conditions. In the experiments of this research, we use a pre-defined path set for each mode based on simple enumeration rules. But in general, it can also be done through heuristics such as column generation. 
\begin{align}
\mathcal{P}^{rs}_{k \mid g(m), t} &= \frac{e^{-\frac{1}{\theta_{m, g(m)}} c^{rs}_{m, g(m), k, t}}}{\sum_{k \in P^{rs}_{m,g(m)}} e^{-\frac{1}{\theta_{m, g(m)}} c^{rs}_{m, g(m), k, t}}} \label{eq-1} \\
\mathcal{P}^{rs}_{g(m) \mid m, t} &= \frac{e^{\frac{\theta_{m, g(m)}}{\theta_m} \cdot IV^{rs}_{m, g(m), t}}}{\sum_{g(m) \in G(m)} e^{\frac{\theta_{m, g(m)}}{\theta_m} \cdot IV^{rs}_{m, g(m), t}}} \\
IV^{rs}_{m, g(m), t} &= \ln \sum_{k \in P^{rs}_{m,g(m)}} e^{-\frac{1}{\theta_{m, g(m)}} c^{rs}_{m, g(m), k, t}} \\
\mathcal{P}^{rs}_{m, t} &= \frac{e^{\theta_m \cdot IV^{rs}_{m, t}}}{\sum_{m \in M} e^{\theta_m \cdot IV^{rs}_{m, t}}} \\
IV^{rs}_{m, t} &= \ln \sum_{g(m) \in G(m)} e^{\frac{\theta_{m, g(m)}}{\theta_m} \cdot IV^{rs}_{m, g(m), t}}\\
h^{rs}_{m,g(m),t} &= q^{rs}_{t} \cdot \mathcal{P}^{rs}_{g(m) \mid m, t} \cdot \mathcal{P}^{rs}_{m, t} \\
f^{rs}_{m,g(m),k,t} &= h^{rs}_{m,g(m),t} \cdot \mathcal{P}^{rs}_{k \mid g(m), t} \\
f^{rs}_{m,g(m),k,t} & \geq 0, \forall k \in P^{rs}_{m,g(m)}, g(m) \in G(m), m \in M, t \in T, (r,s) \in \Omega
\label{eq-1ast}
\end{align}
where \(\theta_m\) and \(\theta_{m,g(m)}\) represent the scale parameter for the first-level nest \(m\) and second-level nest \(g(m)\), respectively, and \(IV_{m,t}^{rs}\) and \(IV_{m,g(m),t}^{rs}\) represent the inclusive value (log-sum term) of the first-level nest \(m\) and second-level nest \(g(m)\), respectively, for O-D pair \((r,s)\) and departure time \(t\).

The multi-modal dynamic equilibrium path flow can be iteratively solved by well-developed methods such as the method of successive averages (MSA), projection-based methods, etc., which are not the focus of the paper. Later, as presented in Section \ref{section - cg}, the iterative process for solving the equilibrium is embedded in the computational graph of the joint dynamic O-D demand and disutility function estimation problem. 

\subsection{Multi-modal dynamic network loading}

The multi-modal dynamic network loading involves the modeling of different types of travel time for cars, buses, metro trains, and passengers. 

\textbf{Travel time of car and bus (with boarding/alighting).} The multi-class traffic flow model proposed in \cite{qian2017modeling}, which models the flow dynamics of multi-class vehicles (e.g., cars and heavy-duty vehicles such as buses), is adopted here. The multi-class cell transmission model (CTM) in \cite{qian2017modeling} is further modified to incorporate passenger boarding and alighting behaviors for buses. Specifically, within the CTM link model, when a bus reaches the cell containing a bus stop on its route, it will stop in that cell if any of the following conditions are met: (1) there are in-vehicle passengers who need to alight at the stop; or (2) there are passengers waiting to board and the number of in-vehicle passengers is below the bus capacity. Vehicle travel times are computed using cumulative arrival and departure curves from the DNL while ensuring the first-in–first-out rule. Separate cumulative curves are maintained for cars and buses. In particular, the cumulative curves for buses are associated with the virtual bus links defined in the bus network in \Cref{sec:multimodal network}. Thus, when a bus reaches or leaves a stop, the corresponding arrival and departure curves on the related virtual bus link increase accordingly.

\textbf{Metro behavior modeling.} Unlike buses, metro trains operate on an exclusive network, and the travel times between metro stations are fixed. Metro stations are modeled as nodes, with links representing direct connections between them. Metro trains are assumed to stop at each station for a minimum fixed dwell time, with additional time added based on boarding and alighting needs. Metro trains also differ physically from buses; for example, they typically have higher capacity and shorter boarding/alighting times per passenger due to multiple doors that allow simultaneous flows.

\textbf{Dwell time.} The dwell time at bus stops or metro stations depends on the boarding and alighting needs. Specifically, buses or metro trains continue boarding passengers waiting in the queue unless capacity is reached or no consecutive boarding demand arises within a loading time interval (5 seconds).

\textbf{Travel time of walking links.} For passengers, cumulative arrival and departure curves are also modeled for the various types of walking links, including: links between origins/destinations and stops, links between physical stops to allow transfers across transit lines, links between parking lots and bus/metro stops for PNR, and the boarding/alighting links between physical and virtual stops. When waiting for buses or metro trains, passengers queue at the downstream end of a walking link, and their waiting time is obtained from the cumulative curves for the corresponding time interval. Walking time is assumed to be proportional to walking distance, based on an assumed walking speed. 

\section{Computational Graph-Based Joint Dynamic O-D Demand and Disutility Function Estimation} \label{sec: dode_cg}
\subsection{Problem formulation}
The objective of the estimation problem is to minimize the discrepancies between the simulated outcomes and the observed data, including dynamic link flows, dynamic link travel times, and passenger boarding and alighting counts at stops in public transit trips.

Denote \(\mathbf{f}_{m,g(m)} = \{f^{rs}_{m,g(m),k,t}, k \in P^{rs}_{m,g(m)}, t \in T, (r,s) \in \Omega\}\) and \(\mathbf{c}_{m,g(m)} = \{c^{rs}_{m,g(m),k,t}, k \in P^{rs}_{m,g(m)}, t \in T, (r,s) \in \Omega\}\). Let \(\mathbf{x}_c\) represent the vector of link flow, i.e., number of cars, for all links, and let \(\mathbf{x}_{pt}\) represent the vector of passenger boarding and alighting counts at all stops for all bus/metro trips during the time horizon of interest. A trip refers to a single run of a bus or metro service, corresponding to the definition of a trip in the General Transit Feed Specification (GTFS).

In practice, observations may be unavailable for some road links, transit stops, or individual bus/metro trips. Let \(\bar{\mathbf{x}}_c\) denote the simulated link flows corresponding to the subset of links for which flow data are observed, and let \(\bar{\mathbf{x}}_{pt}\) denote the simulated boarding and alighting counts at the observed transit stops. The vectors \(\bar{\mathbf{x}}^{'}_c\) and \(\bar{\mathbf{x}}^{'}_{pt}\) represent the actual observed link flows and observed boarding/alighting counts, respectively. Define \(\mathbf{t}_c\) as the vector of travel times across all links, and let \(\bar{\mathbf{t}}_c\) represent the simulated link travel times for the links where travel-time observations exist. The corresponding observed travel-time vector is \(\bar{\mathbf{t}}^{'}_c\). The vectors \(\mathbf{x}_c\) and \(\mathbf{t}_c\) have the dimension of \(|T| \times |A_c|\), where \(T\) is the set of time intervals and \(A_c\) is the complete set of all links. The vector \(\mathbf{x}_{pt}\) has dimension \(2|A_{pt}|\), where \(A_{pt}\) is the set of (stop, trip) pairs, and \(A_{pt}\) is multiplied by 2 to account for both boarding and alighting. Let \(\bar{A}_{c,\mathbf{x}}\), \(\bar{A}_{c,\mathbf{t}}\)  and \(\bar{A}_{pt}\) denote the sets of observed links for car flows, observed links for travel times, and observed transit (stop, trip) pairs, respectively. Accordingly, vectors \(\bar{\mathbf{x}}_c\) and \(\bar{\mathbf{x}}^{'}_c\) have dimensions \(|\bar{A}_{c,\mathbf{x}}|\times|T|\), while \(\bar{\mathbf{t}}_c\) and \(\bar{\mathbf{t}}^{'}_c\) have dimensions \(|\bar{A}_{c,\mathbf{t}}|\times|T|\). The vectors \(\bar{\mathbf{x}}_{pt}\) and \(\bar{\mathbf{x}}^{'}_{pt}\) each have dimension \(2|\bar{A}_{pt}|\).

The following presents the minimization problem for the joint estimation of dynamic O-D demand, denoted as \(\mathbf{q} = \{q^{rs}_t, (r,s) \in \Omega, t \in T\}\), and disutility function (parameters \(\pmb{\beta}\), \(\pmb{\gamma}\) and \(\pmb{\alpha}\)). In the objective function, three types of loss are minimized with weights \(w_i, i=\{1,2,3\}\).
\begin{align}
    \min_{\mathbf{q},\pmb{\beta},\pmb{\gamma},\pmb{\alpha}} \ \ \mathcal{L} = w_1 \|\bar{\mathbf{x}}^{'}_c-\bar{\mathbf{x}}_c\|^2 + w_2 \|\bar{\mathbf{x}}^{'}_{pt}-\bar{\mathbf{x}}_{pt}\|^2 + w_3 \|\bar{\mathbf{t}}^{'}_c-\bar{\mathbf{t}}_c\|^2
    \label{eq-obj}
\end{align}
subject to
\begin{align}
    \big\{\mathbf{x}_c, \mathbf{x}_{pt}, \mathbf{t}_c \big\}  &= \Gamma (\mathbf{f}_{m,g(m)}, g(m) \in G(m), m \in M) \\
    \big\{\mathbf{f}_{m,g(m)}, g(m) \in G(m), m \in M\big\}& = \Delta (\mathbf{q},\pmb{\beta},\pmb{\gamma},\pmb{\alpha}, \Gamma)\\
    \bar{\mathbf{x}}_c & = \mathbf{I}_c\mathbf{x}_c\\
    \bar{\mathbf{x}}_{pt} & = \mathbf{I}_{pt}\mathbf{x}_{pt}\\
    \bar{\mathbf{t}}_c & = \mathbf{J}_c\mathbf{t}_c\\
    \mathbf{q} & \geq 0 \label{eq:last constriant}
\end{align}
where \(\Gamma\) is the mapping of dynamic network loading (DNL), with the input of path flows, it outputs the link flows, link travel times, and boarding and alighting counts. \(\Delta\) is a solver that maps dynamic O-D flow to path flow with the equilibrium given by \Cref{eq-1} - \Cref{eq-1ast} and a set of parameters. With the input \(\mathbf{q},\pmb{\beta},\pmb{\gamma},\pmb{\alpha}\) and the dynamic network loading mapping \(\Gamma\), it outputs the equilibrium path flows. \(\mathbf{I}_c\) is a matrix whose rows represent the observed (link, time interval) pairs in the link flow data and whose columns represent all (link, time interval) pairs. In \(\mathbf{I}_c\), the elements corresponding to the same (link, time interval) pair in both row and column are 1; all other elements are 0. \(\mathbf{I}_{pt}\) is a matrix whose rows represent the observed (stop, trip, boarding/alighting) pairs in the collected data and whose columns represent all (stop, trip, boarding/alighting) pairs. In \(\mathbf{I}_{pt}\), the elements corresponding to the same (stop, trip, boarding/alighting) pair in both row and column are 1; otherwise, they are 0. Similarly, \(\mathbf{J}_c\) is a matrix whose rows represent the observed (link, time interval) pairs in the link travel time data and whose columns represent all (link, time interval) pairs. The elements corresponding to the same (link, time interval) pair in both row and column are 1.

Although the formulated problem intends to estimate the total demand across all modes, the by-mode demand (i.e., mode shares) can also be obtained eventually since all intermediate variables, such as path flows, path costs, etc., will be produced by the simulation. 

In implementation, the weights \(w_1, w_2, w_3\) can be set either to reflect the relative importance of the three types of loss or to normalize their values, given that the magnitudes of different loss components may vary. In the experiments in this study, we assign equal importance to the three loss components and scale their magnitudes by dividing each by its corresponding loss value obtained in the first iteration.

\subsection{Learning using a computational graph}\label{section - cg}
The minimization problem (\ref{eq-obj}) - (\ref{eq:last constriant}) is challenging to solve, since it involves the dynamic network loading mapping that does not have a closed form and it also requires a sub-problem of solving the dynamic equilibrium which itself requires complex iterations. These challenges are further amplified for large-scale networks.
\begin{figure}[ht]
    \centering
    \includegraphics[width=0.95\textwidth]{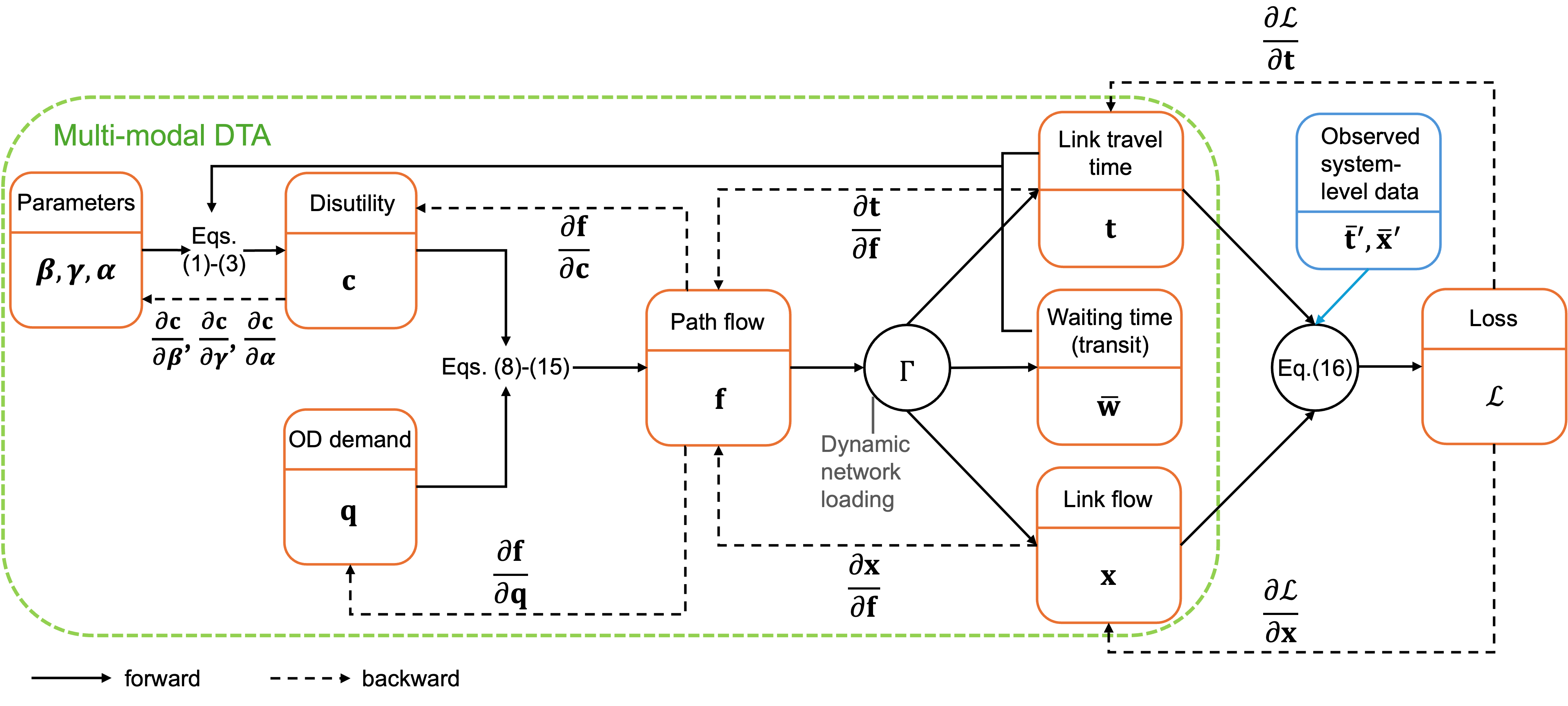}
    \caption{Computational graph representation}
    \label{fig-cg} 
\end{figure}

To address the challenges described above, we leverage a deep learning approach built upon a computational graph representation. \Cref{fig-cg} illustrates the computational graph, where subscripts of the variables are omitted for clarity. A forward–backward process runs iteratively on the computational graph to obtain the solution to the problem. In the forward process, the multi-modal dynamic traffic assignment is solved given any input O–D demand and disutility parameters; this constitutes an embedded inner loop within the forward pass of the computational graph. In the backward process, the gradients of the objective function with respect to the decision variables (\(\mathbf{q},\pmb{\beta},\pmb{\gamma},\pmb{\alpha}\)) are computed step by step based on the chain rule, following the flows shown in the computational graph. In essence, the estimation problem can be cast as a deep learning task that can be solved iteratively using gradient descent methods. Because the problem involves non–closed-form subproblems, automatic differentiation cannot be directly applied. Therefore, all gradients in the backward process are explicitly computed at each iteration and then passed to the embedded optimizers provided by PyTorch, which automates the learning process through mechanisms such as adaptive momentum, learning rate scheduling, and gradient regularization.

The following presents all the gradients involved in the computational graph. The relevant notations are summarized in \Cref{tab:notations}.

\afterpage{
\clearpage
\begin{table}[p]
    \centering
    \caption{Notations used in the forward-backward process}
    \label{tab:notations}  
    \begin{tabular}{l l l l}
        \hline
        Category &Notation & Description & Dimension \\
        \hline
        Demand & \(\mathbf{q}\) & Dynamic O-D demand vector  & \(|\Omega|\times |T|\) \\
        \hline
        Path flow & \(\mathbf{f}_{m,g(m)}\) & Path flow vector for mode \(g(m)\) & \(\sum_{(r,s) \in \Omega} |P^{rs}_{m,g(m)}| \times |T|\)\\
         & \(\mathbf{f}_{c}\) & Path flow vector for car mode & \(\sum_{(r,s) \in \Omega} |P^{rs}_{m,g(m)=car}| \times |T|\)\\
         & \(\mathbf{f}_{pt}\) & \begin{tabular}{@{}l@{}}Path flow vector for mode choices \\ that involve bus or metro\end{tabular} & \begin{tabular}{@{}l@{}}\(\sum_{(r,s) \in \Omega} \sum_{g(m) \in \phi} |P^{rs}_{m,g(m)}| \times |T|\), \\ \(\phi = \{\text{bus,metro,bus+metro}\}\)\end{tabular} \\
         & \(\mathbf{f}_{pnr}\) & \begin{tabular}{@{}l@{}}Path flow vector for mode choices \\ that are PNR trips\end{tabular} & \begin{tabular}{@{}l@{}}\(\sum_{(r,s) \in \Omega} \sum_{g(m) \in \phi} |P^{rs}_{m,g(m)}| \times |T|\), \\ \(\phi = \{\text{car+bus,car+metro,car+bus+metro}\}\)\end{tabular} \\
         & \(\mathbf{f}\) & Path flow vector for all modes & \(\sum_{(r,s) \in \Omega} \sum_{m \in M} \sum_{g(m) \in G(m)} |P^{rs}_{m,g(m)}| \times |T|\)\\
         \hline
         Link flow & \(\mathbf{x}_c\) & Link flow vector for all links & \( |A_c| \times |T|\)\\
         & \(\mathbf{x}_{pt}\) & \begin{tabular}{@{}l@{}}Vector of boarding/alighting count \\ for all (stop,trip) pairs\end{tabular}  & \( 2|A_{pt}|\)\\
         & \(\bar{\mathbf{x}}_c\) & \begin{tabular}{@{}l@{}}Vector of simulated link flow for \\ observed links\end{tabular}  & \( |\bar{A}_{c,\mathbf{x}}| \times |T|\)\\
         & \(\bar{\mathbf{x}}_{pt}\) & \begin{tabular}{@{}l@{}}Vector of simulated boarding/alig- \\ hting for observed (stop,trip) pairs \end{tabular} & \( 2|\bar{A}_{pt}|\)\\
         & \(\bar{\mathbf{x}}^{'}_c\) & Vector of observed link flow & \( |\bar{A}_{c,\mathbf{x}}| \times |T|\)\\
         & \(\bar{\mathbf{x}}_{pt}^{'}\) & \begin{tabular}{@{}l@{}}Vector of observed boarding and \\ alighting count \end{tabular} & \(2|\bar{A}_{pt}|\)\\
         \hline
         \begin{tabular}{@{}l@{}}Link travel \\time \end{tabular}& \(\mathbf{t}_{c}\) & Vector of travel time for all links& \(|A_c| \times |T|\)\\
         & \(\bar{\mathbf{t}}_{c}\) & \begin{tabular}{@{}l@{}}Vector of simulated travel time for \\ all observed links \end{tabular} & \(|\bar{A}_{c,\mathbf{t}}| \times |T|\)\\
         & \(\bar{\mathbf{t}}^{'}_{c}\) & Vector of observed link travel time & \(|\bar{A}_{c,\mathbf{t}}| \times |T|\)\\
         \hline
         Disutility & \(\mathbf{c}_{m,g(m)}\) & Path disutility for mode \(g(m)\) & \(\sum_{(r,s) \in \Omega} |P^{rs}_{m,g(m)}| \times |T|\)\\
         \hline
         \begin{tabular}{@{}l@{}l@{}}Observation \\incidence \\ matrix \end{tabular} & \(\mathbf{I}_{c}\) & \begin{tabular}{@{}l@{}}Observation incidence matrix for \\link flows \end{tabular} & \((|\bar{A}_{c,\mathbf{x}}| \times |T|, |A_{c}| \times |T|)\)\\
         & \(\mathbf{I}_{pt}\) & \begin{tabular}{@{}l@{}}Observation incidence matrix for \\boarding and alighting count \end{tabular} & \((2|\bar{A}_{pt}|, 2|A_{pt}|)\)\\
         & \(\mathbf{J}_{c}\) & \begin{tabular}{@{}l@{}}Observation incidence matrix for \\link travel times \end{tabular} & \((|\bar{A}_{c,\mathbf{t}}| \times |T|, |A_c| \times |T|)\)\\
         \hline 
         DAR matrix & \(\pmb{\rho}_{c}\) & DAR matrix for car mode & \((|A_c| \times |T|, \sum_{(r,s) \in \Omega} |P^{rs}_{m,g(m)=car}| \times |T|)\)\\
         & \(\pmb{\rho}_{pnr}^c\) & \begin{tabular}{@{}l@{}}DAR matrix for mode choices that \\are PNR trips \end{tabular} & \begin{tabular}{@{}l@{}l@{}}\((|A_c| \times |T|\), \\
         \(\sum_{(r,s) \in \Omega} \sum_{g(m) \in \phi} |P^{rs}_{m,g(m)}| \times |T|)\), \\ \(\phi = \{\text{car+bus,car+metro,car+bus+metro}\}\)\end{tabular}\\
         & \(\pmb{\rho}_{pt}\) & \begin{tabular}{@{}l@{}} DAR matrix for mode choices that \\ involve bus or metro\end{tabular} &\begin{tabular}{@{}l@{}l@{}}\((2|A_{pt}|\), \\
         \(\sum_{(r,s) \in \Omega} \sum_{g(m) \in \phi} |P^{rs}_{m,g(m)}| \times |T|)\), \\ \(\phi = \{\text{bus,metro,bus+metro}\}\)\end{tabular}\\
         & \(\pmb{\rho}_{pnr}^{pt}\) & \begin{tabular}{@{}l@{}}DAR matrix for mode choices that \\are PNR trips \end{tabular} & \begin{tabular}{@{}l@{}l@{}}\((2|A_{pt}|\), \\
         \(\sum_{(r,s) \in \Omega} \sum_{g(m) \in \phi} |P^{rs}_{m,g(m)}| \times |T|)\), \\ \(\phi = \{\text{car+bus,car+metro,car+bus+metro}\}\)\end{tabular}\\
        \hline
    \end{tabular}
\end{table}
}

(1) \textit{Derivatives of total loss with respect to path flows}.

For convenience in the derivation, we define the following notations. Denote \(\mathbf{f}_{c}\) as the vector of path flows for second-level mode choice of driving, i.e., \(\mathbf{f}_{c} = \mathbf{f}_{m,g(m)}, g(m)=\text{car}\). Denote \(\mathbf{f}_{pt}\) as the stacked vector of path flows for the second-level mode choices that involve bus or metro travels, i.e.,  \(\mathbf{f}_{pt} = \{\mathbf{f}_{m,g(m)} | g(m) \in \{\text{bus,metro,bus+metro}\}\}\). Denote \(\mathbf{f}_{pnr}\) as the stacked vector of path flows for the second-level mode choices that are PNR trips, i.e.,  \(\mathbf{f}_{pnr} = \{\mathbf{f}_{m,g(m)} | g(m) \in \{\text{car+bus,car+metro,car+bus+metro}\}\}\). Then, we have the following.
\begin{align}
\label{eq: der of loss to f - first}
    \frac{\partial \mathcal{L}}{\partial \mathbf{f}_c} &= w_1\frac{\partial \|\bar{\mathbf{x}}^{'}_c-\bar{\mathbf{x}}_c\|^2}{\partial \mathbf{f}_c} + w_3\frac{\partial \|\bar{\mathbf{t}}^{'}_c-\bar{\mathbf{t}}_c\|^2}{\partial \mathbf{f}_c} \\
    \frac{\partial \|\bar{\mathbf{x}}^{'}_c-\bar{\mathbf{x}}_c\|^2}{\partial \mathbf{f}_c} &= \left(\frac{\partial \mathbf{x}_c}{\partial \mathbf{f}_c}\right)^T \cdot \left(-2 \mathbf{I}_c^T (\bar{\mathbf{x}}^{'}_c - \mathbf{I}_c\mathbf{x}_c)\right) \\
    \frac{\partial \mathbf{x}_c}{\partial \mathbf{f}_c} &= \pmb{\rho}_c \\
    \frac{\partial \|\bar{\mathbf{t}}^{'}_c-\bar{\mathbf{t}}_c\|^2}{\partial \mathbf{f}_c} &= \left(\frac{\partial \mathbf{t}_c}{\partial \mathbf{f}_c}\right)^T \cdot \left(-2 \mathbf{J}_c^T (\bar{\mathbf{t}}^{'}_c - \mathbf{J}_c\mathbf{t}_c)\right) \\
    \frac{\partial \mathbf{t}_c}{\partial \mathbf{f}_c} &= \frac{\partial \mathbf{t}_c}{\partial \mathbf{x}_c} \frac{\partial \mathbf{x}_c}{\partial \mathbf{f}_c} = \frac{\partial \mathbf{t}_c}{\partial \mathbf{x}_c} \cdot \pmb{\rho}_c \\
     \frac{\partial \mathcal{L}}{\partial \mathbf{f}_{pt}} &= w_2\frac{\partial \|\bar{\mathbf{x}}^{'}_{pt}-\bar{\mathbf{x}}_{pt}\|^2}{\partial \mathbf{f}_{pt}} \\
     \frac{\partial \|\bar{\mathbf{x}}^{'}_{pt}-\bar{\mathbf{x}}_{pt}\|^2}{\partial \mathbf{f}_{pt}}  &= \left(\frac{\partial \mathbf{x}_{pt}}{\partial \mathbf{f}_{pt}}\right)^T \cdot \left(-2 \mathbf{I}_{pt}^T (\bar{\mathbf{x}}^{'}_{pt} - \mathbf{I}_{pt}\mathbf{x}_{pt})\right) \\
     \frac{\partial \mathbf{x}_{pt}}{\partial \mathbf{f}_{pt}} &= \pmb{\rho}_{pt} \\
     \frac{\partial \mathcal{L}}{\partial \mathbf{f}_{pnr}} &= w_1\frac{\partial \|\bar{\mathbf{x}}^{'}_c-\bar{\mathbf{x}}_c\|^2}{\partial \mathbf{f}_{pnr}} + w_2\frac{\partial \|\bar{\mathbf{x}}^{'}_{pt}-\bar{\mathbf{x}}_{pt}\|^2}{\partial \mathbf{f}_{pnr}} + w_3\frac{\partial \|\bar{\mathbf{t}}^{'}_c-\bar{\mathbf{t}}_c\|^2}{\partial \mathbf{f}_{pnr}} \\
     \frac{\partial \|\bar{\mathbf{x}}^{'}_c-\bar{\mathbf{x}}_c\|^2}{\partial \mathbf{f}_{pnr}} &= \left(\frac{\partial \mathbf{x}_c}{\partial \mathbf{f}_{pnr}}\right)^T \cdot \left(-2 \mathbf{I}_c^T (\bar{\mathbf{x}}^{'}_c - \mathbf{I}_c\mathbf{x}_c)\right) \\
     \frac{\partial \mathbf{x}_c}{\partial \mathbf{f}_{pnr}} &= \pmb{\rho}^c_{pnr} \\
     \frac{\partial \|\bar{\mathbf{x}}^{'}_{pt}-\bar{\mathbf{x}}_{pt}\|^2}{\partial \mathbf{f}_{pnr}} &= \left(\frac{\partial \mathbf{x}_{pt}}{\partial \mathbf{f}_{pnr}}\right)^T \cdot \left(-2 \mathbf{I}_{pt}^T (\bar{\mathbf{x}}^{'}_{pt} - \mathbf{I}_{pt}\mathbf{x}_{pt})\right) \\
     \frac{\partial \mathbf{x}_{pt}}{\partial \mathbf{f}_{pnr}} &= \pmb{\rho}^{pt}_{pnr} \\
     \frac{\partial \|\bar{\mathbf{t}}^{'}_c-\bar{\mathbf{t}}_c\|^2}{\partial \mathbf{f}_{pnr}} &= \left(\frac{\partial \mathbf{t}_c}{\partial \mathbf{f}_{pnr}}\right)^T \cdot \left(-2 \mathbf{J}_c^T (\bar{\mathbf{t}}^{'}_c - \mathbf{J}_c\mathbf{t}_c)\right) \\
     \frac{\partial \mathbf{t}_c}{\partial \mathbf{f}_{pnr}} &= \frac{\partial \mathbf{t}_c}{\partial \mathbf{x}_c} \cdot \frac{\partial \mathbf{x}_c}{\partial \mathbf{f}_{pnr}} = \frac{\partial \mathbf{t}_c}{\partial \mathbf{x}_c} \cdot \pmb{\rho}^c_{pnr} 
     \label{eq: der of loss to f - last}
\end{align}

\(\pmb{\rho}_c, \pmb{\rho}_{pt}, \pmb{\rho}^c_{pnr}\) and \(\pmb{\rho}^{pt}_{pnr}\) are Dynamic Assignment Ratio (DAR) matrices. The DAR matrices encode how path flows departing at various time intervals are dynamically assigned across network links, and they are produced as part of the DNL model outputs \citep{ma2018estimating}. Specifically, in the context of this research, the elements of  \(\pmb{\rho}_c\) and \(\pmb{\rho}^c_{pnr}\) represent the proportion of traffic flow departing along a given path during a specific time interval that enters a particular link within a certain time interval, for the first-level mode choice of car and PNR, respectively. In \(\pmb{\rho}_c\) and \(\pmb{\rho}^c_{pnr}\), each column represents a unique (path, departure time interval) pair, while each row corresponds to a specific (link, entry time interval) pair. Differently, each row of \(\pmb{\rho}_{pt}\) and \(\pmb{\rho}^{pt}_{pnr}\) corresponds to a specific (stop, trip, boarding/alighting) pair. The elements of these matrices indicate the fraction of passenger flow departing on a particular path and time interval that boards or alights at a particular bus or metro stop on a specific service run.

The above derivation also involves the derivative of link travel time with respect to link flow, i.e., \(\frac{\partial \mathbf{t}_c}{\partial \mathbf{x}_c}\). It is common to assume that the link travel time is differentiable with respect to the incoming link flow, and the derivative is usually obtained based on approximation approaches \citep{Ma2020, lu2013dynamic, qian2012system} since no closed-form solution exists. In this research, the approach in \cite{Ma2020} and \cite{lu2013dynamic} is adopted. The basic idea of this kind of methods is to examine the extra link travel time induced by
a marginal vehicle added to the link. Specifically, \(\frac{\partial \mathbf{t}_c}{\partial \mathbf{x}_c}\) is a square matrix with non-diagonal elements being zero. For elements in diagonal positions, the derivative is zero when the corresponding link is not congested and is the reciprocal of the flow exiting from the head of the link when the corresponding link is congested.

Collecting Eqs. \eqref{eq: der of loss to f - first} - \eqref{eq: der of loss to f - last}, we have
\begin{align}
\label{eq: final der of loss to f - first}
    \frac{\partial \mathcal{L}}{\partial \mathbf{f}_c} &= -2 \left(w_1\pmb{\rho}_c^T \mathbf{I}_c^T (\bar{\mathbf{x}}^{'}_c - \mathbf{I}_c\mathbf{x}_c) + w_3 \pmb{\rho}_c^T\left(\frac{\partial \mathbf{t}_c}{\partial \mathbf{x}_c}\right)^T\mathbf{J}_c^T (\bar{\mathbf{t}}^{'}_c - \mathbf{J}_c\mathbf{t}_c) \right) \\
     \frac{\partial \mathcal{L}}{\partial \mathbf{f}_{pt}} &= -2 w_2 \pmb{\rho}_{pt}^T \mathbf{I}_{pt}^T (\bar{\mathbf{x}}^{'}_{pt} - \mathbf{I}_{pt}\mathbf{x}_{pt}) \\
     \frac{\partial \mathcal{L}}{\partial \mathbf{f}_{pnr}} &= -2 \left( w_1{\pmb{\rho}^c_{pnr}}^T \mathbf{I}_c^T (\bar{\mathbf{x}}^{'}_c - \mathbf{I}_c\mathbf{x}_c) + w_2{\pmb{\rho}^{pt}_{pnr}}^T\mathbf{I}_{pt}^T (\bar{\mathbf{x}}^{'}_{pt} - \mathbf{I}_{pt}\mathbf{x}_{pt}) + w_3 {\pmb{\rho}^c_{pnr}}^T\left(\frac{\partial \mathbf{t}_c}{\partial \mathbf{x}_c}\right)^T \mathbf{J}_c^T (\bar{\mathbf{t}}^{'}_c - \mathbf{J}_c\mathbf{t}_c) \right)
     \label{eq: final der of loss to f - last}
\end{align}

(2) \textit{Derivatives of path flows with respect to O-D demand}.
\begin{align}
    \frac{\partial \mathbf{f}_{c}}{\partial \mathbf{q}} &= \frac{\partial \mathbf{f}_{m,g(m)}}{\partial \mathbf{q}}, g(m)=\text{car} \\
    \frac{\partial \mathbf{f}_{pt}}{\partial \mathbf{q}} &= \Biggl\{\frac{\partial \mathbf{f}_{m,g(m)}}{\partial \mathbf{q}} \Bigg| g(m) \in \{\text{bus,metro,bus+metro}\} \Biggl\}\\
    \frac{\partial \mathbf{f}_{pnr}}{\partial \mathbf{q}} &= \Biggl\{\frac{\partial \mathbf{f}_{m,g(m)}}{\partial \mathbf{q}} \Bigg| g(m) \in \{\text{car+bus,car+metro,car+bus+metro}\} \Biggl\}\\
    \frac{\partial \mathbf{f}_{m,g(m)}}{\partial \mathbf{q}} &= \Biggl\{ \frac{\partial f^{rs}_{m,g(m),k,t}}{\partial q^{r^{'}s^{'}}_{t^{'}}} , k \in P^{rs}_{m,g(m)}; t, t^{'} \in T; (r,s),(r^{'},s^{'}) \in \Omega \Biggl\} \\
    \frac{\partial f^{rs}_{m,g(m),k,t}}{\partial q^{r^{'}s^{'}}_{t^{'}}} &= \begin{cases}
  \frac{\partial f^{rs}_{m,g(m),k,t}}{\partial q^{rs}_{t}} & \text{if } t = t^{'}, (r,s) = (r^{'},s^{'})\\
  0 & \text{o.w.} \\
    \end{cases} \\
    \frac{\partial f^{rs}_{m,g(m),k,t}}{\partial q^{rs}_{t}} &= \mathcal{P}^{rs}_{m,t} \cdot \mathcal{P}^{rs}_{g(m)|m,t} \cdot \mathcal{P}^{rs}_{k|g(m),t}
\end{align}

(3) \textit{Derivatives of path flows with respect to parameters \(\boldsymbol{\beta}\), \(\boldsymbol{\gamma}\), and \(\boldsymbol{\alpha}\)}.

Let \(\mathbf{f}\) be a stacked vector of path flows \(\mathbf{f}_{c}\), \(\mathbf{f}_{pt}\), and \(\mathbf{f}_{pnr}\). Let \(\mathbf{c}\) be a stacked vector of path disutilities \(\mathbf{c}_{m,g(m)}, g(m)\in G(m), m \in M\). We have
\begin{align}
    \frac{\partial \mathbf{f}}{\partial \pmb{\beta}} = \frac{\partial \mathbf{f}}{\partial \mathbf{c}} \frac{\partial \mathbf{c}}{\partial \pmb{\beta}}, \quad \frac{\partial \mathbf{f}}{\partial \pmb{\gamma}} = \frac{\partial \mathbf{f}}{\partial \mathbf{c}} \frac{\partial \mathbf{c}}{\partial \pmb{\gamma}}, \quad \frac{\partial \mathbf{f}}{\partial \pmb{\alpha}} = \frac{\partial \mathbf{f}}{\partial \mathbf{c}} \frac{\partial \mathbf{c}}{\partial \pmb{\alpha}}
\end{align}

For the values in \(\frac{\partial \mathbf{c}}{\partial \pmb{\beta}}, \frac{\partial \mathbf{c}}{\partial \pmb{\gamma}}\), and \(\frac{\partial \mathbf{c}}{\partial \pmb{\alpha}}\), we have
\begin{align}
    \frac{\partial c^{rs}_{m,g(m),k,t}}{\partial \beta_{1,car}} &= \begin{cases}
  w^{rs}_{k,t,car} & \text{if }  g(m) \in \{\text{car, car+bus, car+metro, car+bus+metro}\} \\
  0 & \text{o.w.} \\
    \end{cases} \\
    \frac{\partial c^{rs}_{m,g(m),k,t}}{\partial \beta_{1,bus}} &= w^{rs}_{k,t,bus} \cdot \xi_{bus}^{g(m)}\\
    \frac{\partial c^{rs}_{m,g(m),k,t}}{\partial \beta_{1,metro}} &= w^{rs}_{k,t,metro} \cdot \xi_{metro}^{g(m)}\\
    \frac{\partial c^{rs}_{m,g(m),k,t}}{\partial \beta_{2}} &= \begin{cases}
  \tau_k^{rs} & \text{if }  g(m) = \text{car} \\
  \delta_k^{rs} & \text{if }  g(m) \in \{\text{bus, metro, bus+metro}\}\\
  \tau_k^{rs} + \delta_k^{rs}  & \text{if }  g(m) \in \{\text{car+bus, car+metro, car+bus+metro}\}
    \end{cases} \\
    \frac{\partial c^{rs}_{m,g(m),k,t}}{\partial \beta_{3,bus}} &= \bar{w}^{rs}_{k,t,bus} \cdot \xi_{bus}^{g(m)}\\
    \frac{\partial c^{rs}_{m,g(m),k,t}}{\partial \beta_{3,metro}} &= \bar{w}^{rs}_{k,t,metro} \cdot \xi_{metro}^{g(m)}\\
    \frac{\partial c^{rs}_{m,g(m),k,t}}{\partial \beta_{4}} &= \begin{cases}
  0 & \text{if }  g(m) = \text{car} \\
  \tilde{w}^{rs}_{k,t} & \text{o.w.}\\
  \end{cases}\\
  \frac{\partial c^{rs}_{m,g(m),k,t}}{\partial \gamma_{1, m^{'},g(m)^{'}}} &= \begin{cases}
  I^{rs} & \text{if }  (m,g(m)) = (m^{'},g(m)^{'}) \\
  0 & \text{o.w.}\\
  \end{cases}\\
  \frac{\partial c^{rs}_{m,g(m),k,t}}{\partial \gamma_{2, m^{'},g(m)^{'}}} &= \begin{cases}
  J^{r} & \text{if }  (m,g(m)) = (m^{'},g(m)^{'}) \\
  0 & \text{o.w.}\\
  \end{cases}\\
  \frac{\partial c^{rs}_{m,g(m),k,t}}{\partial \gamma_{3, m^{'},g(m)^{'}}} &= \begin{cases}
  J^{s} & \text{if }  (m,g(m)) = (m^{'},g(m)^{'}) \\
  0 & \text{o.w.}\\
  \end{cases}\\
  \frac{\partial c^{rs}_{m,g(m),k,t}}{\partial \alpha_{m^{'},g(m)^{'}}} &= \begin{cases}
  1 & \text{if }  (m,g(m)) = (m^{'},g(m)^{'}) \\
  0 & \text{o.w.}\\
  \end{cases}
\end{align}

For values \( \frac{\partial f^{rs}_{m,g(m),k,t}}{\partial c^{r's'}_{m',g'(m'),k',t'}} \) in \(\frac{\partial \mathbf{f}}{\partial \mathbf{c}}\), we have the following four cases of non-zero values. To save space, the detailed derivations of Eqs. \eqref{eq: f/c case1} - \eqref{eq: f/c case4} are presented in Appendix A.

\textbf{a}. The derivative of the path flow  \(f^{rs}_{m,g(m),k,t}\) with respect to the disutility of the same path \(c^{rs}_{m,g(m),k,t}\).
\begin{align}
    \frac{\partial f^{rs}_{m,g(m),k,t}}{\partial c^{rs}_{m,g(m),k,t}} &= f^{rs}_{m,g(m),k,t} \cdot \Bigg[
    -\frac{1}{\theta_{m,g(m)}} (1 - \mathcal{P}^{rs}_{k|g(m),t})
    - \frac{1}{\theta_m} \mathcal{P}^{rs}_{k|g(m),t}(1 - \mathcal{P}^{rs}_{g(m)|m,t}) \nonumber \\ 
    &\quad - \mathcal{P}^{rs}_{k|g(m),t} \mathcal{P}^{rs}_{g(m)|m,t} (1 - \mathcal{P}^{rs}_{m,t})
    \Bigg]
    \label{eq: f/c case1}
\end{align}

\textbf{b}. The derivative of the path flow \(f^{rs}_{m,g(m),k,t}\) with respect to the disutility of a different path in the same second-level mode \(c^{rs}_{m,g(m),k',t}, k \neq k'\).
\begin{align}
    \frac{\partial f^{rs}_{m,g(m),k,t}}{\partial c^{rs}_{m,g(m),k',t}} &= f^{rs}_{m,g(m),k,t} \cdot \Bigg[
    \frac{1}{\theta_{m,g(m)}} \mathcal{P}^{rs}_{k'|g(m),t}
    - \frac{1}{\theta_m} \mathcal{P}^{rs}_{k'|g(m),t}(1 - \mathcal{P}^{rs}_{g(m)|m,t}) \nonumber \\
    &\quad - \mathcal{P}^{rs}_{k'|g(m),t} \mathcal{P}^{rs}_{g(m)|m,t} (1 - \mathcal{P}^{rs}_{m,t})
    \Bigg]
    \label{eq: f/c case2}
\end{align}

\textbf{c}. The derivative of the path flow \(f^{rs}_{m,g(m),k,t}\) with respect to the disutility of a different path in a different second-level mode but same first-level mode \(c^{rs}_{m,g'(m),k',t}, k \neq k', g(m) \neq g'(m)\).
\begin{align}
    \frac{\partial f^{rs}_{m,g(m),k,t}}{\partial c^{rs}_{m,g'(m),k',t}} &= f^{rs}_{m,g(m),k,t} \cdot \mathcal{P}^{rs}_{g'(m)|m,t} \cdot \mathcal{P}^{rs}_{k'|g'(m),t} \cdot \Bigg(
    \frac{1}{\theta_m}  - 1 + \mathcal{P}^{rs}_{m,t}\Bigg)
    \label{eq: f/c case3}
\end{align}

\textbf{d}. The derivative of the path flow \(f^{rs}_{m,g(m),k,t}\) with respect to the disutility of a different path in an entirely different first-level mode \(c^{rs}_{m',g'(m'),k',t}, k \neq k', g(m) \neq g'(m'), m \neq m' \).
\begin{align}
    \frac{\partial f^{rs}_{m,g(m),k,t}}{\partial c^{rs}_{m',g'(m'),k',t}} &= f^{rs}_{m,g(m),k,t} \cdot \mathcal{P}^{rs}_{m',t} \cdot \mathcal{P}^{rs}_{g'(m')|m',t} \cdot \mathcal{P}^{rs}_{k'|g'(m'),t}
    \label{eq: f/c case4}
\end{align}

So far, based on the above derivation, we have obtained all the gradients used in the computational graph. The gradients of the total loss with respect to all estimation variables, i.e., \(\mathbf{q},\pmb{\beta},\pmb{\gamma},\pmb{\alpha}\), can be obtained based on chain rule:
\begin{align*}
    \frac{\partial \mathcal{L}}{\partial \mathbf{q}} &= \Big(\frac{\partial \mathbf{f}}{\partial \mathbf{q}}\Big)^T \frac{\partial \mathcal{L}}{\partial \mathbf{f}}, \quad \frac{\partial \mathcal{L}}{\partial \pmb{\beta}} = \Big(\frac{\partial \mathbf{f}}{\partial \pmb{\beta}}\Big)^T \frac{\partial \mathcal{L}}{\partial \mathbf{f}}, \quad \frac{\partial \mathcal{L}}{\partial \pmb{\gamma}} = \Big(\frac{\partial \mathbf{f}}{\partial \pmb{\gamma}}\Big)^T\frac{\partial \mathcal{L}}{\partial \mathbf{f}}, \quad \frac{\partial \mathcal{L}}{\partial \pmb{\alpha}} = \Big(\frac{\partial \mathbf{f}}{\partial \pmb{\alpha}}\Big)^T\frac{\partial \mathcal{L}}{\partial \mathbf{f}} \\
    \frac{\partial \mathcal{L}}{\partial \mathbf{f}} &= \Biggl\{\Big(\frac{\partial \mathcal{L}}{\partial \mathbf{f}_c}\Big)^T, \Big(\frac{\partial \mathcal{L}}{\partial \mathbf{f}_{pt}}\Big)^T, \Big(\frac{\partial \mathcal{L}}{\partial \mathbf{f}_{pnr}}\Big)^T  \Biggl\}^T, \quad \frac{\partial \mathbf{f}}{\partial \mathbf{q}} = \Biggl\{\Big(\frac{\partial \mathbf{f}_c}{\partial \mathbf{q}}\Big)^T, \Big(\frac{\partial \mathbf{f}_{pt}}{\partial \mathbf{q}}\Big)^T, \Big(\frac{\partial \mathbf{f}_{pnr}}{\partial \mathbf{q}}\Big)^T  \Biggl\}^T
\end{align*}

Leveraging the computational graph and learning technologies, we can address the following challenges: (a) the ability to handle large-scale networks and high-dimensional variables/parameters that need to be estimated; (b) the capability to conduct sophisticated dynamic modeling of travel time (i.e., dynamic network loading) to account for congestion effects; and (c) the ability to work with and leverage any combination of multi-source data (even with missing data).

\section{Hypothesis Testing} \label{sec: hypo}
A hypothesis testing framework is proposed to enable the evaluation of the statistical significance of estimated parameters (\(\boldsymbol{\beta}\), \(\boldsymbol{\gamma}\), and \(\boldsymbol{\alpha}\)) for potential variable selection and insightful behavioral analysis in real-world implementation. In this research, parameters (\(\boldsymbol{\beta}\), \(\boldsymbol{\gamma}\), and \(\boldsymbol{\alpha}\)) are estimated by minimizing the discrepancy, measured by the sum of squared residuals, between observed real-world data and the simulated outputs generated by a complex nonlinear model. Therefore, the estimation is a nonlinear regression problem based on nonlinear least squares (NLS), or more broadly, an M-estimation problem \citep{davidson1993estimation}. The hypothesis testing in the context of this research is challenging due to the following reasons: (1) Non-closed-form traffic dynamics: the estimation problem involves a dynamic network loading model that lacks a closed-form solution; (2) High dimensionality: the number of parameters in general is high-dimensional, depending on the multiple factors from multi-source data and the scaling by the number of alternatives if they are alternative-specific; (3) The complexity resulting from the joint estimation of dynamic O-D demand and the parameters.  

To address these challenges, we employ an approach based on Wald principle \citep{wald1943tests, davidson1993estimation}. First, since the estimation involves a complex simulation model with non-closed-form formulations, we apply the Wald principle that does not require a closed-form likelihood formulation. Second, due to the high dimensionality of parameters to be estimated, we apply this approach that conducts joint executions, i.e., obtaining the statistics based on covariance matrix, for multiple parameters altogether instead of testing them one by one to save computational time. 

Let \( \boldsymbol{\theta} = \{ \boldsymbol{\beta}\), \(\boldsymbol{\gamma}\), \(\boldsymbol{\alpha}\} \) be a parameter vector for easy presentation in this section. \( \boldsymbol{\theta} \) is estimated by solving the optimization problem:
\[
\hat{\boldsymbol{\theta}} = \arg\min_{\boldsymbol{\theta}} \mathcal{L}(\boldsymbol{\theta})
\]
where \(\mathcal{L}(\boldsymbol{\theta})\) is a scalar loss function measuring the discrepancy (sum of squared residuals) between the observed data and the simulation output, as defined in \Cref{eq-obj}.

The goal is to test the following hypothesis for each component \( \theta_j \in \boldsymbol{\theta}\), and we usually have \(\theta_{j,0} =0 \).
\[
H_0: \theta_j = \theta_{j,0} \quad \text{vs.} \quad H_1: \theta_j \neq \theta_{j,0}
\]

Under mild conditions, the NLS-estimator \( \hat{\boldsymbol{\theta}} \) is asymptotically normal \citep{davidson1993estimation}, i.e.,
\[
\sqrt{n} (\hat{\boldsymbol{\theta}} - \boldsymbol{\theta}^-) \xrightarrow{d} \mathcal{N}(0, \Sigma)
\]
where \(n\) is the effective number of independent observations being matched in the loss function and \(\boldsymbol{\theta}^-\) represents the true value of parameter. This condition represents that as \(n \rightarrow \infty\), the distribution of \(\sqrt{n} (\hat{\boldsymbol{\theta}} - \boldsymbol{\theta}^-)\) converges to a multivariate normal distribution with mean zero and finite covariance matrix. The Wald principle \citep{wald1943tests} is to construct a test
statistic based on unrestricted parameter estimates and an estimate of the unrestricted covariance matrix. Since our hypothesis involves just one restriction, i.e., \(\theta_j = \theta_{j,0}\), we calculate the following pseudo-t statistic for hypothesis testing with respect to \(\theta_j, \forall j\), which is asymptotically distributed as \(N(0,1)\) \citep{wald1943tests, davidson1993estimation}.
\[
z_j = \frac{\hat{\theta}_j - \theta_{j,0}}{\text{SE}(\hat{\theta}_j)}
\]
where the standard error \( \text{SE}(\hat{\theta}_j) = \sqrt{\left[ \text{Var}(\hat{\boldsymbol{\theta}}) \right]_{jj}} \). 

According to \cite{davidson1993estimation}, the estimated covariance matrix of \(\hat{\boldsymbol{\theta}}\), \(\text{Var}(\hat{\boldsymbol{\theta}})\), can be approximated by
\[ 
\text{Var}(\hat{\boldsymbol{\theta}}) \approx \hat{\sigma}^2 \left(\boldsymbol{J}^T\boldsymbol{J}\right)^{-1}  \quad \text{with} \quad \hat{\sigma}^2 = \frac{\mathcal{L}(\hat{\boldsymbol{\theta}})}{n - p}, \boldsymbol{J}_{i,j} = \frac{\partial y_{i}(\boldsymbol{\theta})}{\partial \theta_j} \bigg|_{\theta_j = \hat{\theta}_j}
\]
where \(\boldsymbol{J}\) is the Jacobian matrix of simulated variables \(\boldsymbol{y}\) (i.e., \(\bar{\mathbf{x}}_c, \bar{\mathbf{x}}_{pt}, \bar{\mathbf{t}}_c\) in this research) in the loss function with respect to the estimator \(\hat{\boldsymbol{\theta}}\). \(\hat{\sigma}^2\) is an estimate of the residual variance where \( n \) is the effective number of independent observations in the loss, and \( p \) is the number of estimated parameters.

Then the p-value with respect to \(\theta_j\) is given by
\[
p_j = 2 \times (1 - \Phi(|z_j|))
\]
where \( \Phi(\cdot) \) is the standard normal cumulative distribution function. The null hypothesis is rejected at level \(\alpha\) if \(p_j < \alpha\), indicating that \(\theta_j \) is significantly different from \(\theta_{j,0}\).

A \( (1 - \alpha)\% \) confidence interval for \( \hat{\theta}_j \) is
\[
\left[ \hat{\theta}_j - z_{\alpha/2} \cdot \text{SE}(\hat{\theta}_j),\; \hat{\theta}_j + z_{\alpha/2} \cdot \text{SE}(\hat{\theta}_j) \right]
\]
where \(z_{\alpha/2}\) is the critical value of the standard normal distribution. 

For this research, the Jacobian matrix \(\boldsymbol{J}\) can be obtained based on the derivations in \Cref{section - cg}. Note that since we have three different types of loss in the loss function, and each of them have different statistical characteristics and different number of observations, the hypothesis test is done using the three types of loss separately. Those parameters that are tested as non-significant for all the three cases will be considered non-significant overall.

For real-world tasks, we can iteratively select the variables in the disutility functions based on the results of hypothesis tests. That is, keep the variables that are statistically significant and remove those that are not. After updating the disutility functions, we can conduct the estimation again to obtain a new set of estimated parameters and then run another round of hypothesis testing. This process repeats iteratively until a set of variables/parameters that are all statistically significant is reached. The complete procedure is shown in Algorithm 1. 

\begin{algorithm}[!htbp] 
\SetKwInOut{KwIn}{Input}
\SetKwInOut{KwOut}{Output}

\KwIn{
    Observed data \(\bar{\mathbf{x}}^{'}, \bar{\mathbf{t}}^{'}\) (subscripts omitted); Initial parameter values \( \boldsymbol{\theta}^{(0)} = \{ \boldsymbol{\beta}^{(0)}, \boldsymbol{\gamma}^{(0)}, \boldsymbol{\alpha}^{(0)}\}\) and O-D demand \(\mathbf{q}^{(0)}\) ; Loss function \(\mathcal{L}(\mathbf{q}, \boldsymbol{\beta}, \boldsymbol{\gamma}, \boldsymbol{\alpha})\); Number of maximum iterations \(K\); Rejection threshold \(\alpha\)
}

\KwOut{
    Updated disutility functions with estimated parameters \(\boldsymbol{\theta}^* = \{\boldsymbol{\beta}^*, \boldsymbol{\gamma}^*, \boldsymbol{\alpha}^*\}\); Confidence intervals
}

\For{\(k = 1\) \textbf{to} \(K\)}{
    \textbf{Conduct the joint estimation of O-D demand and disutility functions:}
    
    \Indp
    Conduct the joint estimation by minimizing the loss in \Cref{eq-obj} based on the computational graph approach, with the initial \(\mathbf{q}^{k-1}, \boldsymbol{\theta}^{k-1}\), and get the estimated \(\mathbf{q}^k, \boldsymbol{\theta}^k\).\par
    \Indm

    \textbf{Compute the covariance matrix of \(\boldsymbol{\theta}^k\):}

    \Indp
    Compute \(\hat{\sigma}^2\) and the Jacobian matrix \(\boldsymbol{J}\): 
    \[
    \hat{\sigma}^2 = \frac{\mathcal{L}(\mathbf{q}^k, \boldsymbol{\theta}^k)}{n - p(\boldsymbol{\theta}^k)},  \hat{\sigma}^2 = \frac{\mathcal{L}(\hat{\boldsymbol{\theta}})}{n - p}, \boldsymbol{J}_{i,j} = \frac{\partial y_{i}(\boldsymbol{\theta})}{\partial \theta_j} \bigg|_{\theta_j = \theta_j^k}
    \] \\
    Compute the covariance matrix: 
    \[
    \text{Var}(\boldsymbol{\theta}^k) \approx \hat{\sigma}^2\left(\boldsymbol{J}^T\boldsymbol{J}\right)^{-1}
    \] \\
    Pseudo inverse can be utilized if needed.
    \par
    \Indm

    \textbf{Compute the standard error and test statistic for \(\theta_j^k, \forall \theta_j^k \in \boldsymbol{\theta}^k\):}
    \[
    \text{SE}(\theta_j^k) = \sqrt{\left[ \text{Var}(\boldsymbol{\theta}^k) \right]_{jj}}, z_j = \frac{\theta_j^k - \theta_{j,0}}{\text{SE}(\theta_j^k)}
    \]

    \textbf{Compute the p-value of \(\theta_j^k , \forall \theta_j^k \in \boldsymbol{\theta}^k\):}
    \[
    p_j = 2 \times (1 - \Phi(|z_j|))
    \]

    \If{\(p_j < \alpha, \forall \theta^k_j \in \boldsymbol{\theta}^k\)}{
    \(\boldsymbol{\theta}^* \gets \boldsymbol{\theta}^k\); \\
    \textbf{Break}
    } 
    \vspace{1em}
    \textbf{Update the disutility functions:}

    \Indp
    Remove any \(\theta_j\) with \(p_j > \alpha\) and the associated variables in the disutility functions, and get the updated \(\boldsymbol{\theta}^{k}\) \par
    \Indm

    \If{k = K}{
    \(\boldsymbol{\theta}^* \gets \boldsymbol{\theta}^k\)
    } 
}

\textbf{Compute the \( (1 - \alpha)\% \) confidence interval for \(\theta_j^*, \forall \theta_j^* \in \boldsymbol{\theta}^* \):} \\
\hspace{1.2em} \(
\left[ \theta_j^* - z_{\alpha/2} \cdot \text{SE}(\theta_j^*),\; \theta_j^* + z_{\alpha/2} \cdot \text{SE}(\theta_j^*) \right]
\)

\caption{Hypothesis Testing of Parameters and Variable Selection of Disutility Functions}
\end{algorithm}

\section{Numerical Experiments} \label{sec: experiments}
\begin{figure}[ht]
    \centering
    \includegraphics[width=0.55\textwidth]{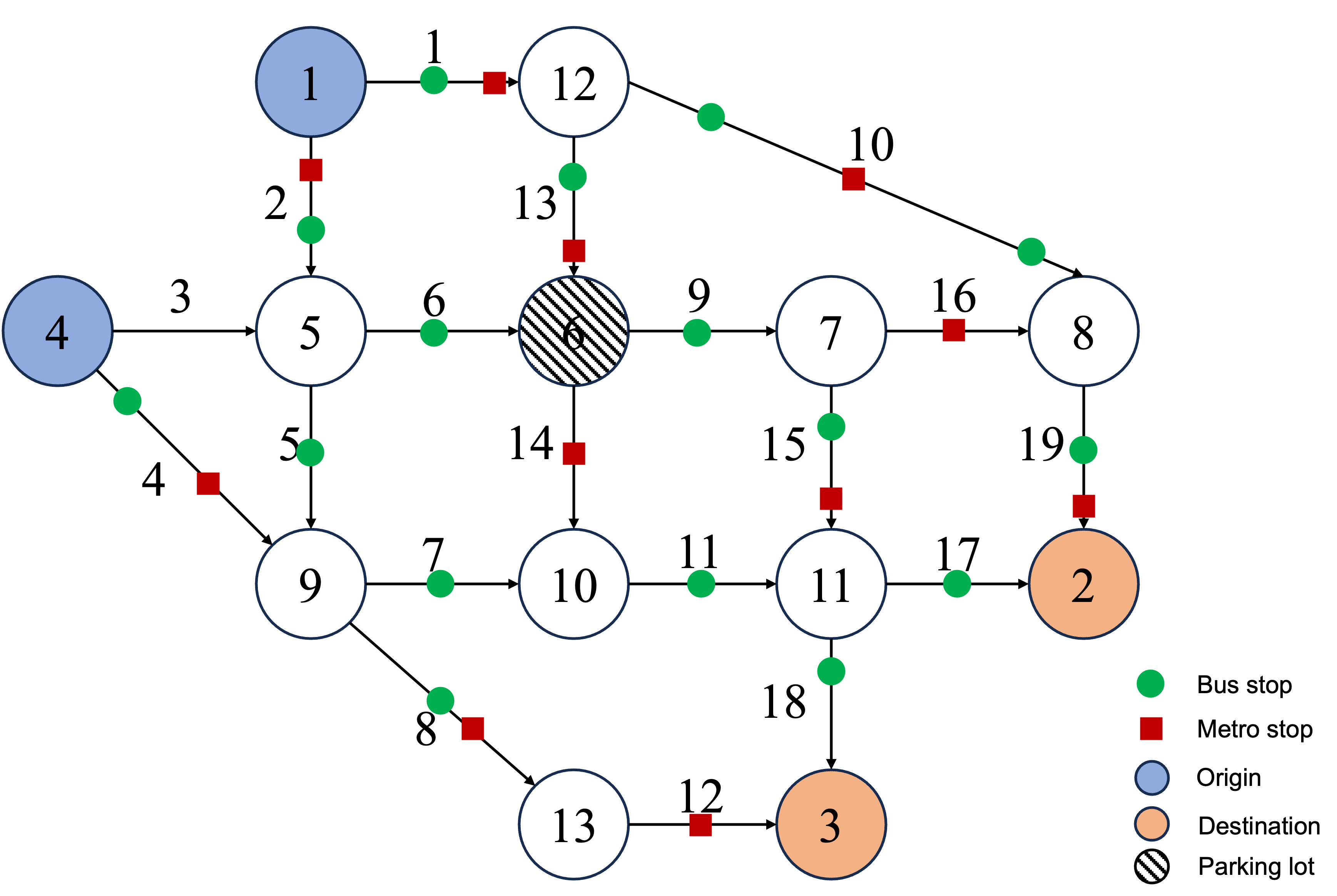}
    \caption{Nguyen-Dupuis network}
    \label{fig-ND} 
\end{figure}

\begin{figure}[ht]
    \centering
    \includegraphics[width=0.8\textwidth]{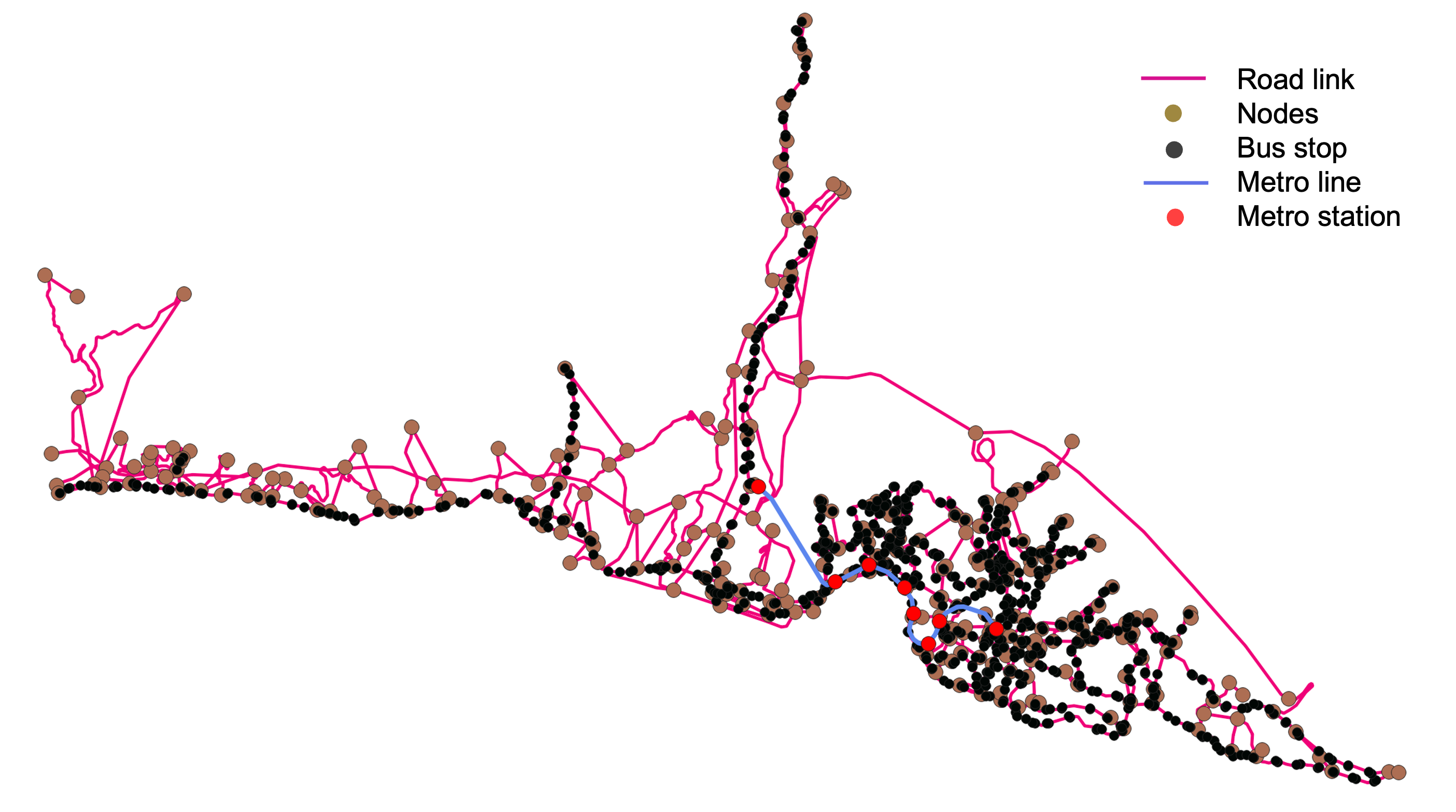}
    \caption{Genoa network}
    \label{fig-genova} 
\end{figure}

The numerical experiments are conducted on a small network (Nguyen-Dupuis network) and a real-world network (Genoa, Italy). For each network, we first build the multi-modal network following the framework proposed in \Cref{sec:multimodal network}. The two networks are shown in \Cref{fig-ND} and \Cref{fig-genova}, respectively. The network specifics with multi-modal elements are shown in \Cref{tab - network specifics}. For both networks, the dynamic O-D demand is estimated at a 15-minute temporal resolution, and the dynamic network loading frequency is every 5 seconds. 

\begin{table}[ht]
    \centering
    \caption{Multi-modal networks specification}
    \label{tab - network specifics}  
    \begin{tabular}{l l l}
        \hline
        Name & Value \\
        \hline
         & Nguyen-Dupuis & Genoa \\
        \hline
        Number of nodes & 13 & 224 \\
        Number of links & 19 & 573 \\
        Number of OD pairs & 4 & 3906 \\
        Number of bus lines & 6 & 56 \\
        Number of metro lines & 3 & 2 \\
        Number of physical bus/metro stops & 27 & 761 \\
        Number of virtual bus/metro stops & 39 & 1170 \\
        Number of (stop, trip) pairs & 1170 & 21797 \\
        Number of walking links & 220 & 12360 \\
        Number of paths (driving) & 25 & 11501 \\
        Number of paths (transit) & 177 & 8016 \\
        Number of paths (PNR) & 84 & 7959 \\
        Number of parking lots & 1 & 4 \\
        Number of estimating variables & 50 & 62530 \\
        \hline
    \end{tabular}
\end{table}

Due to the unavailability of real-world data, we use assumed ground-truth data (e.g., link flows, link travel times, and bus/metro boarding and alighting counts at stops) which are synthetic outputs generated by the simulation using an assumed ground-truth dynamic O–D demand and a set of assumed disutility parameters. For the Genoa network, we have obtained a reference O–D demand dataset from Hitachi Rail, which is used to generate the ground-truth traffic dynamics; the details will be introduced later.

Through a set of experiments on the two networks, we intend to explore the following questions.

\begin{itemize}
    \item Can the proposed framework gradually learn to fit the observed data? This will be evaluated by the Goodness of Fit for the three types of data (dynamic link flow, link travel time, and boarding/alighting counts at stops). A higher goodness of fit provides confidence in using the estimated (or calibrated) models for further what-if analysis and policy design.
    \item Are the three types of loss gradually decreasing over the iterations during the estimation process? This will be evaluated by examining the change in the normalized Mean Squared Error (MSE) between the observed data and the simulation outcomes over iterations. A decreasing trend would indicate that the proposed framework is actively learning, i.e., moving from a state of higher loss to lower loss, by adjusting the dynamic O–D demand and disutility parameters.
    \item Although this is impossible for real-world implementation, we can compare the estimated dynamic O-D demand and parameters with the assumed ``ground truth'' because synthetic data are used in the experiments. This helps us understand whether the proposed framework can ``learn'' the assumed truth, even though the solution to this complex problem is very likely non-unique and the converged solution may easily fall into local optimal. 
    \item We intend to explore numerical evidence on the non-uniqueness of the solution (dynamic O-D demand and parameter values). Here, the term ``non-uniqueness" is more on a practical perspective beyond the mathematical perspective. That is, for real-world tasks where we never know the ground-truth O-D demand, are there non-unique sets of O-D demand that can lead to satisfactory results on data matching? How far are they from the ground truth? And what are the insights for real-world implementation given such non-uniqueness?
    \item Can the framework handle large-scale networks with thousands of O–D pairs, city-level public transit services, and hundreds of thousands of estimation variables on a highly granular temporal basis (e.g., 15 minutes)?
\end{itemize}

We use the Nguyen-Dupuis network for detailed analysis and discussions, and the Genoa network to validate the scalability of the proposed framework. The results presented in this section, together with the discussions in \Cref{sec-Discussions} will answer the questions mentioned above.

\subsection{Experiments on the Nguyen-Dupuis network}
The link characteristics such as capacity, speed limit, jam density for Nguyen-Dupuis network, the parking fees, and the social demographic features such as income and population density for different O-D pairs are manually made up. They are shown in Appendix B. The simulation time window for the traffic dynamics is one hour.

\subsubsection{Goodness of fit and loss change}
In this experiment, the initial O-D demand is obtained by randomly adding/reducing up to 50\% noise, i.e., [-50\%,50\%] random noise, to the ``ground truth'' O-D demand, and the parameters of disutility functions are initialized by adding 10\% noise to the ``ground truth'' values.

\begin{figure}[ht]
    \centering
    \includegraphics[width=1\textwidth]{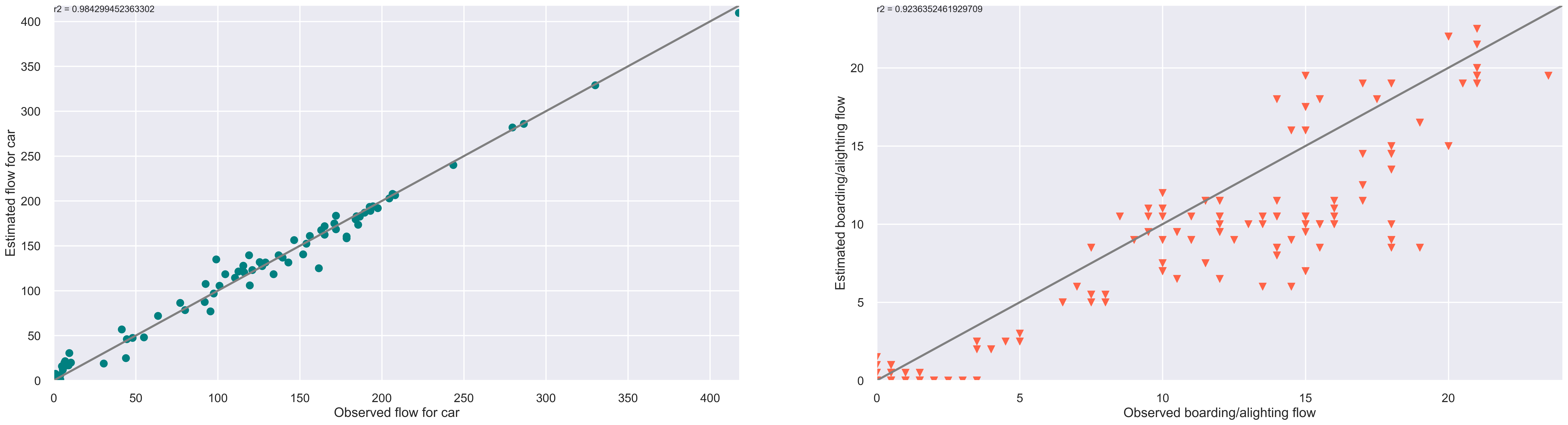}
    \caption{Goodness of fit of link flow (unit: count per 15 minutes) and trip-stop-level bus/metro boarding/alighting counts (Nguyen-Dupuis)}
    \label{fig-ND results - flow fitting} 
\end{figure}

\begin{figure}[ht]
    \centering
    \includegraphics[width=0.68\textwidth]{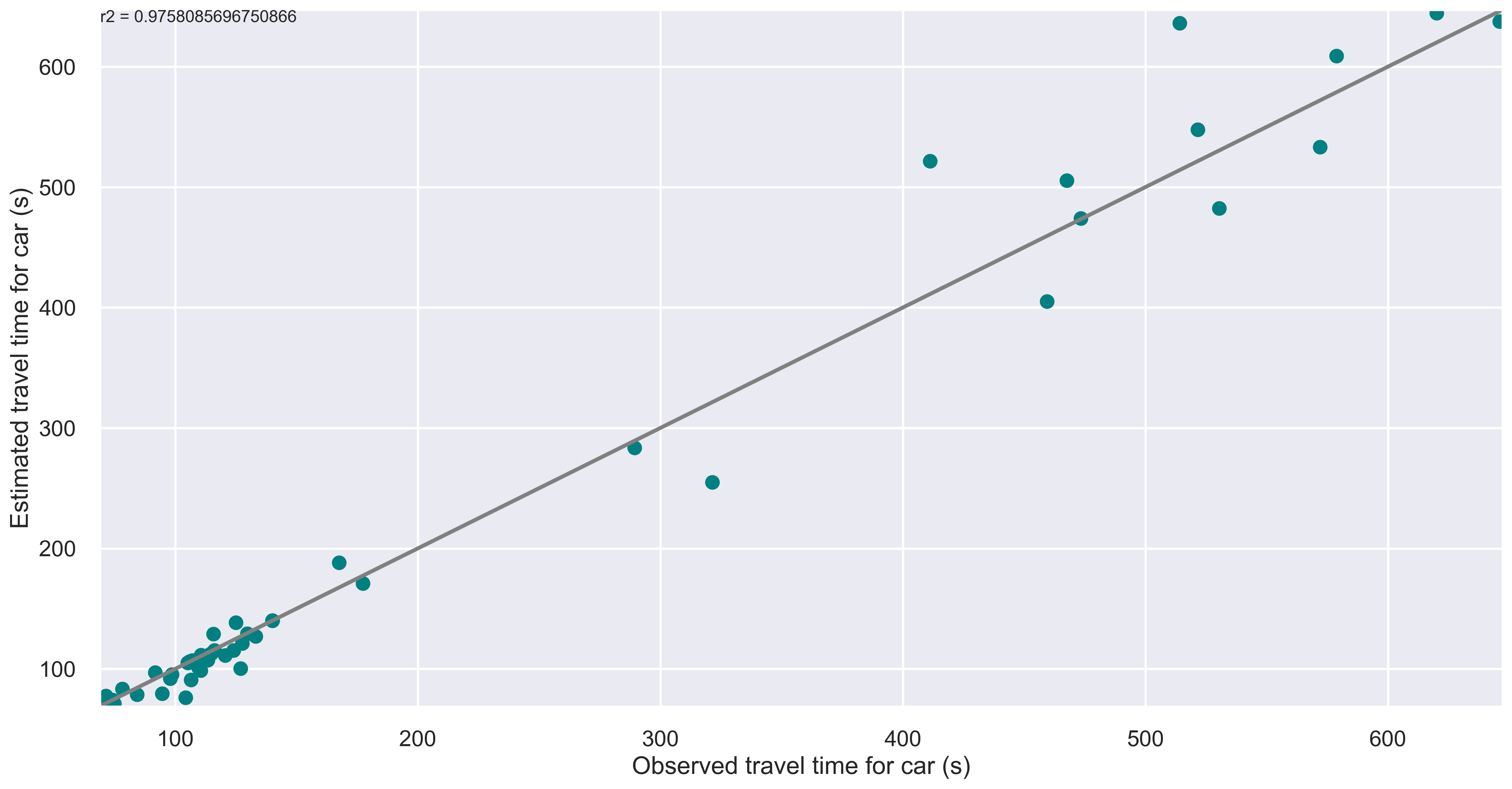}
    \caption{Goodness of fit of dynamic link travel time for Nguyen-Dupuis network}
    \label{fig-ND results - tt fitting} 
\end{figure}

\Cref{fig-ND results - flow fitting} and \Cref{fig-ND results - tt fitting} show the goodness-of-fit results for link flow, trip-stop-level bus/metro boarding and alighting counts, and link travel time. For link flow and link travel time, each dot represents the value for a specific link at a 15-minute interval. For the boarding and alighting counts, each dot represents the number of passengers who board or alight at a particular stop on a specific bus or metro trip. The goodness of fit is measured by the coefficient of determination, denoted \(R^2\). From the figures, the \(R^2\) values are 0.984, 0.924, and 0.976, for link flow, boarding/alighting count, and link travel time, respectively, indicating that the multi-source datasets are matched reasonably well through our learning framework. 

Note that matching boarding and alighting counts at the stop-trip level is more challenging than matching link flow and travel time data, because the former is mathematically more discrete and defined on a much more granular basis. Additional discussion on this point is provided in \Cref{sec-Discussions}.
\begin{figure}[ht]
    \centering
    \includegraphics[width=0.68\textwidth]{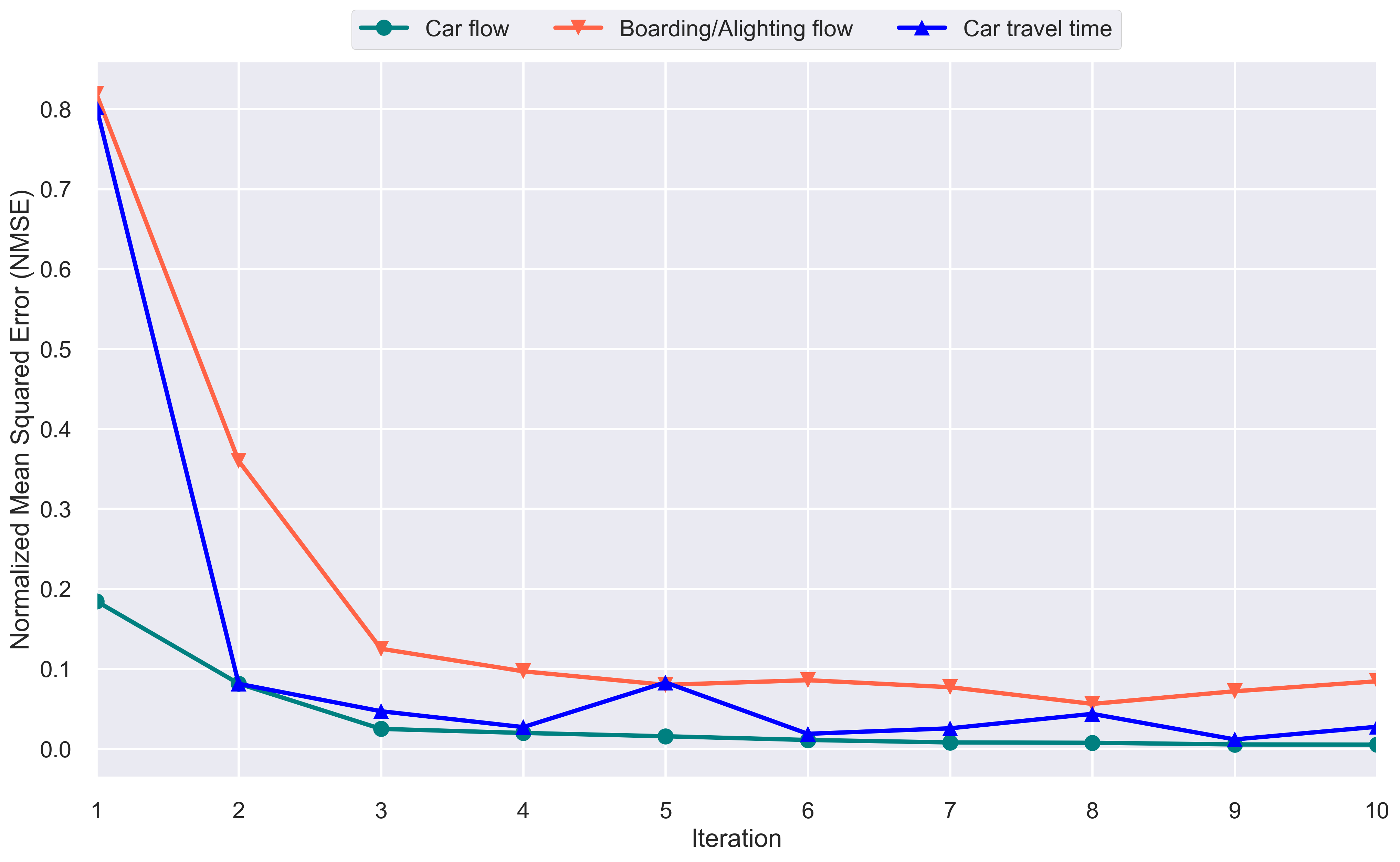}
    \caption{Change of normalized MSE over iterations (Nguyen-Dupuis)}
    \label{fig-ND results - NMSE} 
\end{figure}
\begin{figure}[ht]
    \centering
    \includegraphics[width=0.99\textwidth]{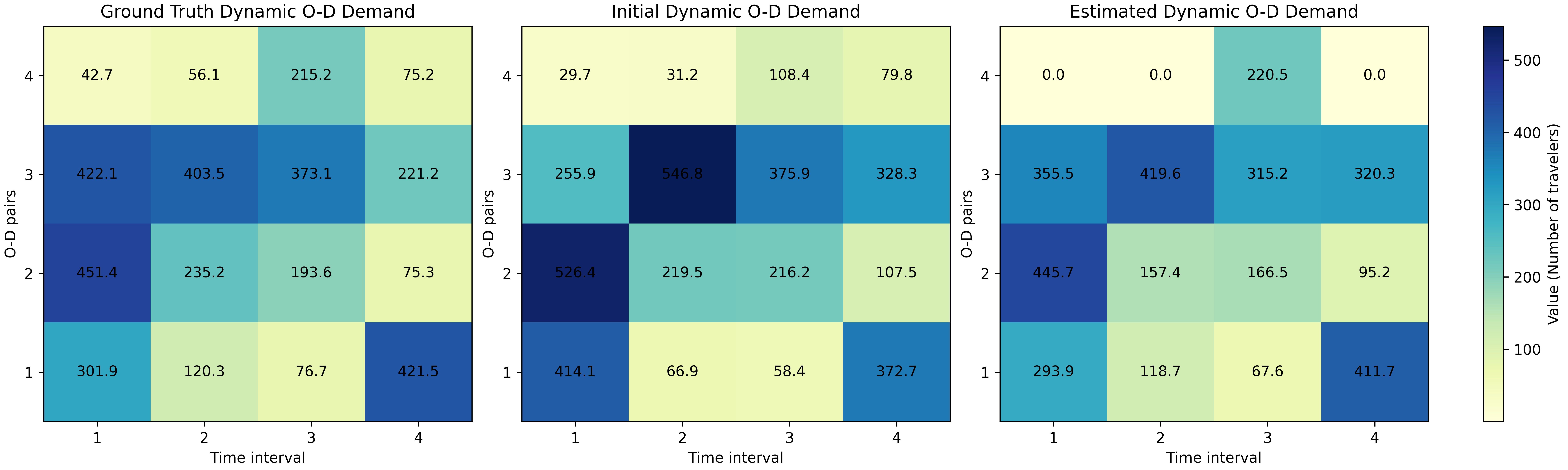}
    \caption{Comparison of ground truth, initial, and estimated dynamic O-D demand (Nguyen-Dupuis)}
    \label{fig-ND results - OD comparison} 
\end{figure}

\Cref{fig-ND results - NMSE} shows the change of the normalized Mean Squared Error (MSE) of the three types of loss. The normalized MSE is computed by MSE divided by mean square of the observed values. We observe that the normalized MSE for each loss component steadily decreases across iterations, indicating effective learning.

\Cref{fig-ND results - OD comparison} shows the comparison of ground truth, initial, and estimated dynamic O-D demand for Nguyen-Dupuis network. We can see that the estimated dynamic O–D demand is overall much closer to the ground truth compared to the initial values, which validates the effectiveness of the estimation. But does a set of O–D demands always need to be close to the ``ground truth" in order to fit the observed traffic dynamics well? Later in \Cref{sec-Discussions}, we provide additional experiments and further discussion on this question.
\FloatBarrier

\subsubsection{Disutility parameter estimation and hypothesis testing}
This subsection focuses on the results of estimated parameters in the disutility functions and hypothesis testing. There are 34 parameters to be estimated in the disutility function, which is a relatively large number for hypothesis testing, especially given the context where the estimation problem is already highly complex. 

For the results of parameter estimation and hypothesis testing, we focus on three aspects. First, whether the learned parameters have the same sign (+ or -) as the ground truth. The consistent sign implies that the model correctly learns the positive or negative impact of factors in the disutility function, which is valuable and crucial for analyzing the economic impacts of different factors in real-world tasks. Second, whether the statistical significance is consistent with the assumed ground truth. Consistent results indicate that hypothesis testing works well in identifying the significant factors that affect travelers' choices, which is also critical for behavior interpretations in real-world tasks. Third, whether the parameters are learned towards the right direction. Learning towards the right direction indicates that the backward process of the computational graph works towards getting closer to the ground-truth values. However, as mentioned before, the solution is very likely non-unique for this complex problem. We expect that the results will help indicate whether non-uniqueness exists—that is, whether there may be sets of parameters that do not closely align with the assumed ``ground truth" yet still yield good data fitting performance and decreasing loss.

\Cref{tab - hypo ND} shows the results for Nguyen-Dupuis network experiments. First, we can observe that all the learned parameters have a consistent sign with the assumed ground truth. Second, since we added 10\% noise to the ground truth to generate the initial points, the result with an absolute percentage error less than 0.1 would imply that the learning is towards a right direction. From the results, we have 9 out of 27 non-zero parameters in the incorrect direction. This implies that even though some parameters do not get closer to the assumed ``ground truth" we can still obtain satisfactory results in the data fitting and loss change (as presented before). \Cref{sec-Discussions} will have more discussions related to this. 

For this experiment, we adopt the significance level of 0.05. In the assumed ground truth, we set \(\gamma_3\), the coefficient of population density of the destination zone, for all the second-level modes except the car as nonsignificant, and also set the \(\gamma_{1,car}\), the coefficient of income of mode car as nonsignificant. The testing correctly turns out 3 of these parameters as nonsignificant, while having incorrect results for the other ones. Most results for all other parameters that are assumed to be significant are all correctly consistent with the ground truth, except the \(\beta_{3,metro}\) (with p-value 0.115). For real-world tasks, we suggest iteratively removing those nonsignificant variables based on the estimation results to gradually shrink the parameter set to finally obtain the disutility function with all parameters significant and updated estimated values. Some nonsignificant parameters may not be tested out in the initial runs. We omitted these iterative steps since the experiments here are not on real-world data, and the experiments are for demonstration purposes only. 

\begin{table}[!htbp]
    \centering
    \caption{Disutility parameter estimation and hypothesis testing results (Nguyen-Dupuis)}
    \label{tab - hypo ND}  
    \begin{tabular}{l l l l l l l l}
        \hline
        Parameter & \begin{tabular}{@{}l@{}}Ground truth \\ value\end{tabular} & \begin{tabular}{@{}l@{}}Estimated \\ value\end{tabular} & \begin{tabular}{@{}l@{}l@{}}Absolute \\ percentage \\ error \end{tabular} & \begin{tabular}{@{}l@{}}Right \\ direction\end{tabular} & \begin{tabular}{@{}l@{}}Right \\ sign\end{tabular} & \begin{tabular}{@{}l@{}}Consistent \\ significance\end{tabular}\\
        \hline
        \(\beta_{1,car}\) & 1 & 1.130 & 0.131 & N & Y & Y \\
        \(\beta_{1,bus}\) & 1.2 & 1.323 & 0.100 & - & Y & Y \\
        \(\beta_{1,metro}\) & 0.8 & 0.857 & 0.071 & Y & Y & Y \\
        \(\beta_{2}\) & 1.5 & 1.673 & 0.115 & N & Y & Y \\
        \(\beta_{3,bus}\) & 1.5 & 1.65& 0.100 & - & Y & Y \\
        \(\beta_{3,metro}\) & 0.8 & 0.855 & 0.068 & Y & Y & N \\
        \(\beta_{4}\) & 1.5 & 1.628 & 0.085 & Y & Y & Y \\
        \(\gamma_{1,car}\) & 0 & 0.07 & - & - & - & Y \\
        \(\gamma_{1,bus}\) & 1 & 1.1 & 0.1 & - & Y & Y \\
        \(\gamma_{1,metro}\) & 0.8 & 0.859 & 0.074 & Y & Y & Y \\
        \(\gamma_{1,bus+metro}\) & 1 & 1.1 & 0.100 & - & Y & Y \\
        \(\gamma_{1,car+bus}\) & 0.3 & 0.333 & 0.109 & N & Y & Y \\
        \(\gamma_{1,car+metro}\) & 0.2 & 0.196 & 0.020 & Y & Y & Y \\
        \(\gamma_{1,car+bus+metro}\) & 0.3 & 0.330 & 0.100 & - & Y & Y \\
        \(\gamma_{2,car}\) & 0.1 & 0.134 & 0.336 & N & Y & Y \\
        \(\gamma_{2,bus}\) & -0.3 & -0.330 & 0.100 & - & Y & Y \\
        \(\gamma_{2,metro}\) & -0.4 & -0.461 & 0.153 & N & Y & Y \\
        \(\gamma_{2,bus+metro}\) & -0.3 & -0.330 & 0.100 & - & Y & Y \\
        \(\gamma_{2,car+bus}\) & 0.1 & 0.113 & 0.127 & N & Y & Y \\
        \(\gamma_{2,car+metro}\) & 0.1 & 0.085 & 0.152 & N & Y & Y \\
        \(\gamma_{2,car+bus+metro}\) & 0.1 & 0.110 & 0.100 & - & Y & Y \\
        \(\gamma_{3,car}\) & 0.3 & 0.133 & 0.555 & N & Y & Y \\
        \(\gamma_{3,bus}\) & 0 & 0.05 & - & - & - & N \\
        \(\gamma_{3,metro}\) & 0 & 0.029 & - & - & - & Y \\
        \(\gamma_{3,bus+metro}\) & 0 & 0.05 & - & - & - & N \\
        \(\gamma_{3,car+bus}\) & 0 & 0.053 & - & - & - & N \\
        \(\gamma_{3,car+metro}\) & 0 & 0.026 & - & - & - & Y \\
        \(\gamma_{3,car+bus+metro}\) & 0 & 0.05 & - & - & - & N \\
        \(\alpha_{bus}\) & 2 & 2.2 & 0.100 & - & Y & Y \\
        \(\alpha_{metro}\) & 1 & 1.079 & 0.079 & Y & Y & Y \\
        \(\alpha_{bus+metro}\) & 1.5 & 1.650 & 0.100 & - & Y & Y \\
        \(\alpha_{car+bus}\) & 1 & 1.106 & 0.106 & N & Y & Y \\
        \(\alpha_{car+metro}\) & 0.5 & 0.527 & 0.051 & Y & Y & Y \\
        \(\alpha_{car+bus+metro}\) & 1 & 1.100 & 0.100 & - & Y & Y \\
        \hline
        * \footnotesize{Y: yes; N: no} & & & & & & 
    \end{tabular}
\end{table}

\subsection{Experiments on the Genoa network}
The physical attributes of the Genoa network, such as link capacity, speed limits, and the number of lanes, are obtained from OpenStreetMap and Google Maps. The sociodemographic features (e.g., income and population density) for different O–D pairs, the GTFS data, and a reference O–D demand dataset are provided by Hitachi Rail. We use this O–D demand dataset from Hitachi Rail as the reference basis for the ground truth, and additional data processing and scaling are performed to generate the final ground truth data. Note that in real-world settings, it is very likely that the observed data do not cover all links or all public transit stops/trips. The proposed formulation \Cref{eq-obj}–\Cref{eq:last constriant} and the computational framework are designed to function effectively even in the presence of missing data.

The purpose of the experiments on the Genoa network is to assess scalability and computational burden on large-scale networks. In the experiment, the simulation time window is extended to 4 hours to represent a typical peak-period duration. Although there is no appropriate benchmark in the literature, we aim to provide a general sense of the computational resources and time needed by the proposed framework to demonstrate its scalability.

All experiments are conducted on a MacBook Pro with an M3 chip and 16 GB of memory. \Cref{tab - computational time} summarizes the computational time required for the experiments on the two networks. The computational time for running one round of DNL is 0.13 seconds for the Nguyen–Dupuis network and 12.44 seconds for the Genoa network, indicating high computational efficiency of the program. 

The computational time per estimation iteration, which includes both the time for multiple DTA iterations and the learning time for estimation, is largely affected by the number of iterations needed to solve the DTA in the forward process of the computational graph. With a convergence threshold of \(10^{-3}\), we allow a maximum of 500 DTA iterations for the Nguyen–Dupuis network. As shown, it takes only 103 seconds per estimation iteration. For the Genoa network, to save time, we limit the maximum number of DTA iterations to 5 (which can yield satisfactory results, as presented later), since the main purpose here is to provide a general sense of the computational burden and one can imagine that if we increase the number of DTA iterations the computational time per iteration for the estimation will increase accordingly. Overall, recall the network scale provided in \Cref{tab - network specifics}, these results demonstrate the high computational efficiency and scalability of the proposed framework. 

\begin{table}[!htbp]
    \centering
    \caption{Computational time of the experiments (Unit: seconds)}
    \label{tab - computational time}  
    \begin{tabular}{l l l l l l l l}
        \hline
        Network & \begin{tabular}{@{}l@{}}Simulation \\ time window \end{tabular} & \begin{tabular}{@{}l@{}}One round \\ of DNL \end{tabular}  &  \begin{tabular}{@{}l@{}l@{}} One iteration of  \\ estimation (including \\ the DTA loop)\end{tabular} & \begin{tabular}{@{}l@{}}Max iterations \\ of the DTA \end{tabular} & \begin{tabular}{@{}l@{}l@{}}Convergence  \\ threshold \\ used in DTA \end{tabular}\\
        \hline
        Nguyen-Dupuis & 1 hour & 0.13 & 103 & 500 & \(10^{-3}\)  \\
        Genoa & 4 hours & 12.44 & 251.1 & 5 & \(10^{-3}\) \\
        \hline
    \end{tabular}
\end{table}

\subsubsection{Goodness of fit and loss change}
In this experiment, the initial O-D demand is obtained by randomly adding/reducing up to 30\% noise, i.e., [-30\%,30\%] random noise, to the ground truth O-D demand, and the parameters of disutility functions are initialized by adding 10\% noise to the ground truth values.

\begin{figure}[ht]
    \centering
    \includegraphics[width=1\textwidth]{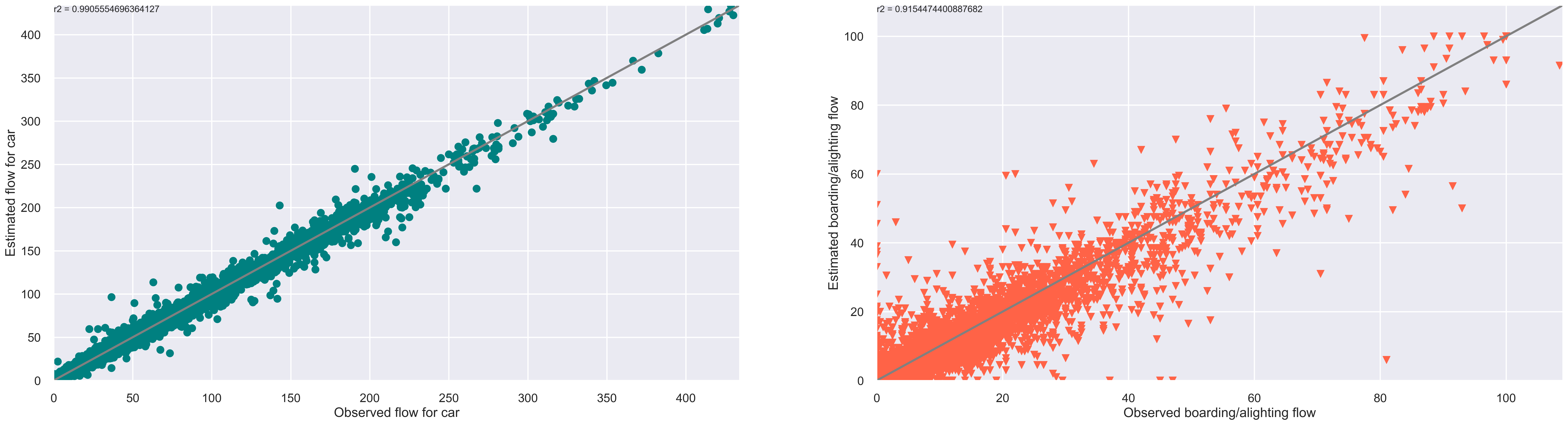}
    \caption{Goodness of fit of link flow (unit: count per 15 minutes) and trip-stop-level bus/metro boarding/alighting counts (Genoa)}
    \label{fig-Genoa results - flow fitting} 
\end{figure}

\begin{figure}[ht]
    \centering
    \includegraphics[width=0.7\textwidth]{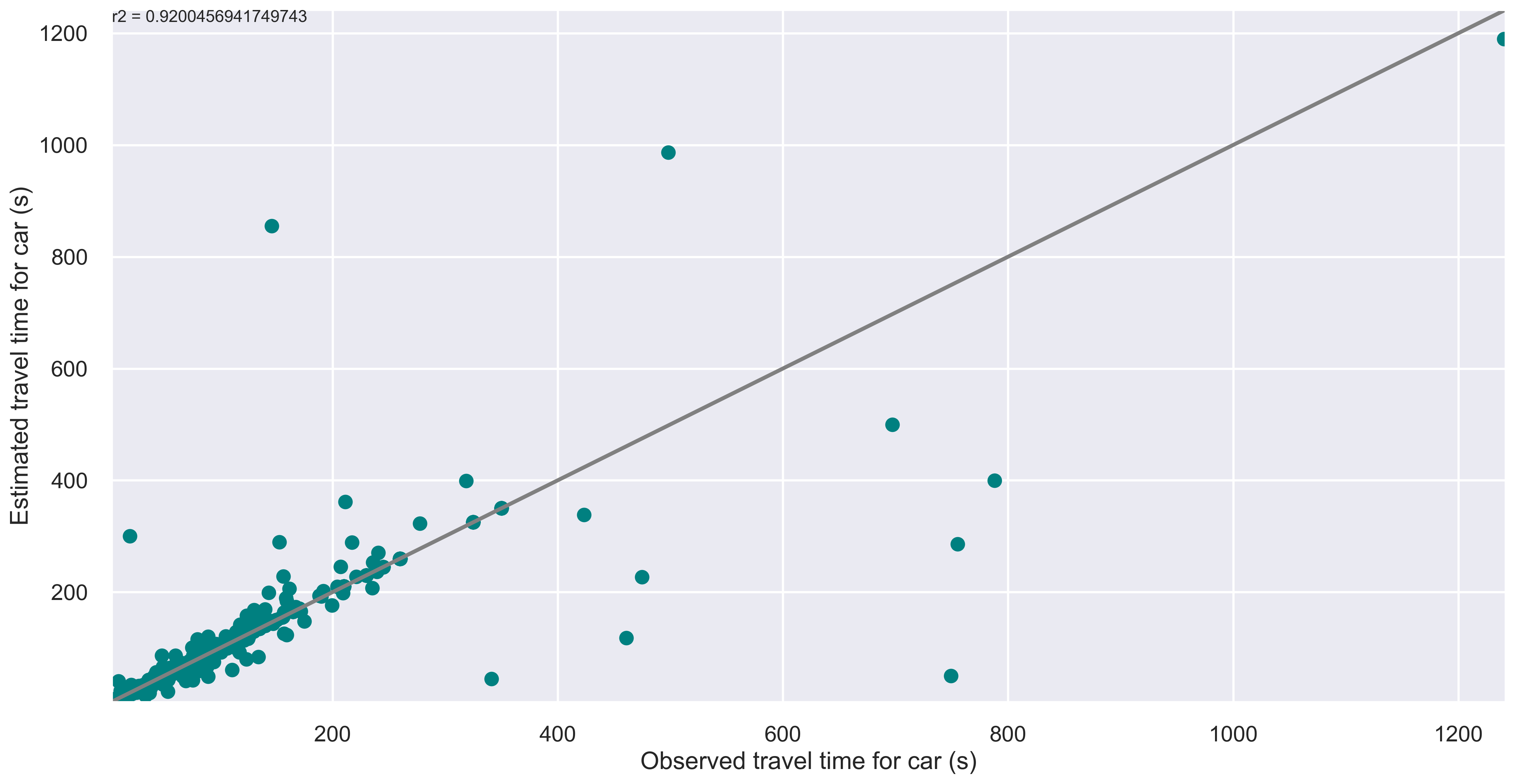}
    \caption{Goodness of fit of dynamic link travel time (Genoa)}
    \label{fig-Genoa results - tt fitting} 
\end{figure}

\Cref{fig-Genoa results - flow fitting} and \Cref{fig-Genoa results - tt fitting} show the goodness-of-fit results for link flow, trip-stop-level bus/metro boarding/alighting counts, and link travel time. For link flow and link travel time, each dot represents the value of a specific link at a 15-minute interval. For the boarding/alighting counts, each dot represents the boarding/alighting number of passengers at a specific stop on a specific bus/metro trip. Similarly to previous experiments, the goodness of fit is measured by the coefficient of determination, denoted \(R^2\). From the figures, the \(R^2\) values are 0.991, 0.915, and 0.920 for link flow, boarding/alighting counts, and link travel time, respectively, indicating that the multi-source datasets are matched reasonably well through our learning framework. 


\begin{figure}[ht]
    \centering
    \includegraphics[width=0.7\textwidth]{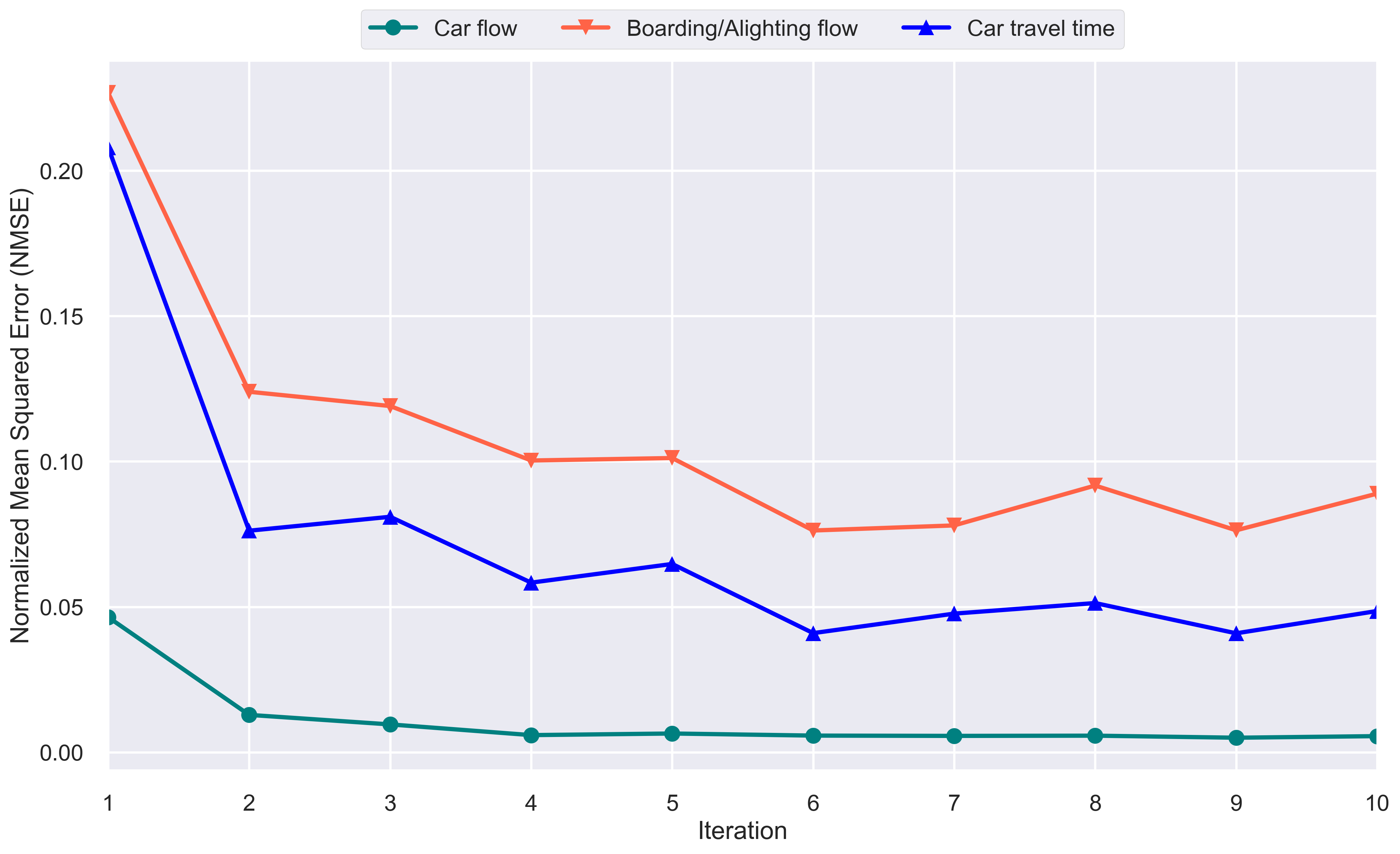}
    \caption{Change of normalized MSE over iterations (Genoa)}
    \label{fig-Genoa results - NMSE} 
\end{figure}

\begin{figure}[!htbp]
    \centering
    \includegraphics[width=0.7\textwidth]{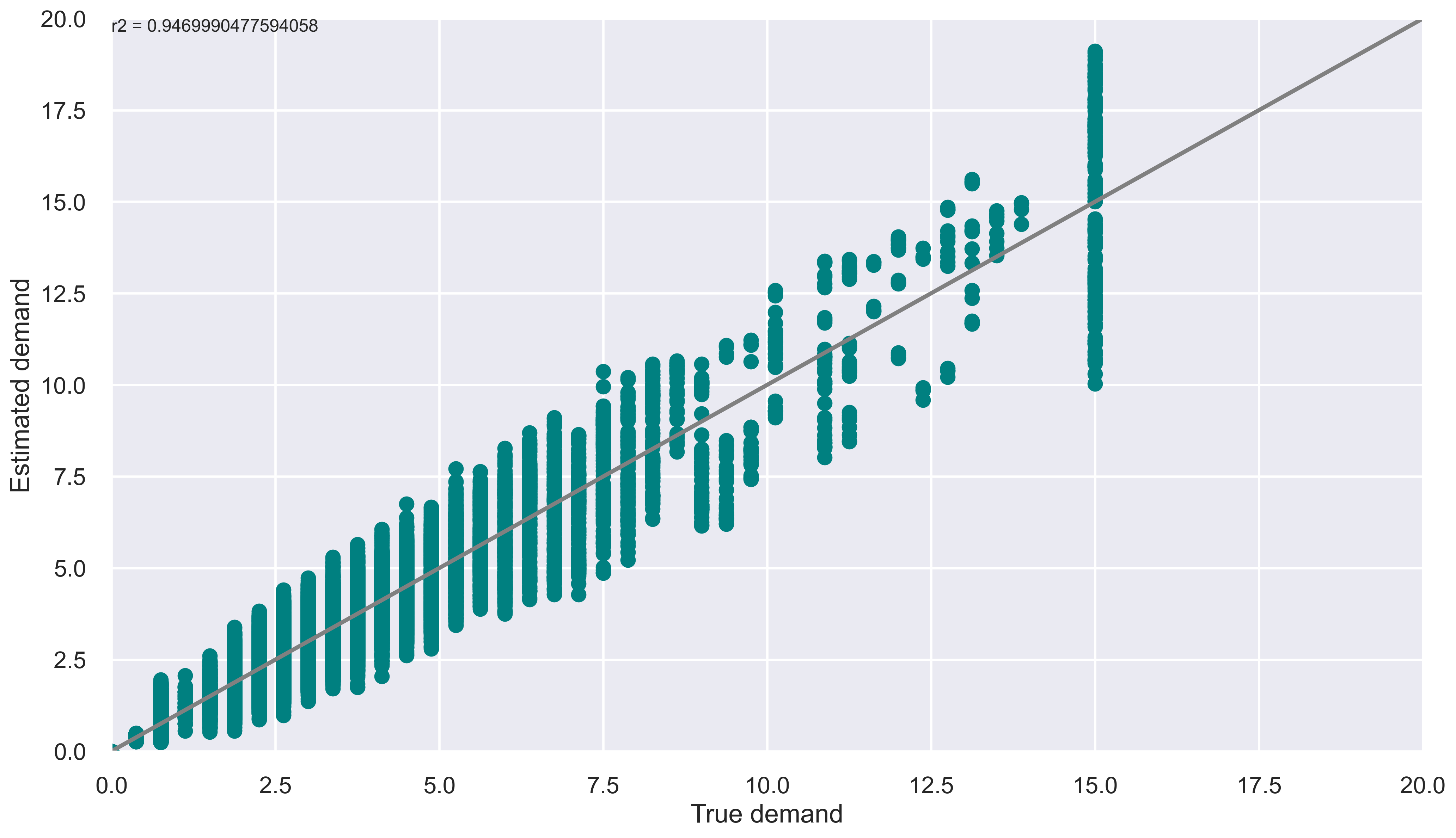}
    \caption{Comparison of ground-truth and estimated (15-minute granularity) O-D demand (Genoa)}
    \label{fig-Genoa results - OD comparison} 
\end{figure}

\Cref{fig-Genoa results - NMSE} shows the change in the normalized MSE of the three types of loss in the loss function. We can see that the normalized MSE gradually decreases for each type of loss over iterations. \Cref{fig-Genoa results - OD comparison} shows the comparison between the ground-truth and estimated dynamic O-D demand for the Genoa network. The estimated dynamic O-D demand matches the ground truth reasonably well (\(R^2 = 0.947\)). 

The results in this subsection demonstrate the capability of the proposed framework when applied to large-scale networks, both technically and computationally.

\subsubsection{Disutility parameter estimation and hypothesis testing}
This subsection focuses on the results of the estimated parameters in the disutility functions and the corresponding hypothesis testing. The same metrics used in the Nguyen–Dupuis network experiments are applied here.

\Cref{tab - hypo Genoa} presents the results for the Genoa network experiments. From the results, we observe that all learned parameters have signs consistent with the assumed ground truth, and most (24 out of 28) nonzero parameters are in the correct direction. We set \(\gamma_3\), the coefficient of population density of the destination zone, to be insignificant for all second-level modes except for car; however, the testing results indicate that these parameters are significant. Additionally, \(\gamma_{1,car+metro}\), the coefficient of income for the car+metro mode, is assumed to be significant in ground truth but is tested as nonsignificant.   

\begin{table}[!htbp]
    \centering
    \caption{Disutility parameter estimation and hypothesis testing results (Genoa)}
    \label{tab - hypo Genoa}  
    \begin{tabular}{l l l l l l l l}
        \hline
        Parameter & \begin{tabular}{@{}l@{}}Ground truth \\ value\end{tabular} & \begin{tabular}{@{}l@{}}Estimated \\ value\end{tabular} & \begin{tabular}{@{}l@{}l@{}}Absolute \\ percentage \\ error \end{tabular} & \begin{tabular}{@{}l@{}}Right \\ direction\end{tabular} & \begin{tabular}{@{}l@{}}Right \\ sign\end{tabular} & \begin{tabular}{@{}l@{}}Consistent \\ significance\end{tabular}\\
        \hline
        \(\beta_{1,car}\) & 0.5 & 0.560 & 0.120 & N & Y & Y \\
        \(\beta_{1,bus}\) & 0.4 & 0.424 & 0.061 & Y & Y & Y \\
        \(\beta_{1,metro}\) & 0.3 & 0.312 & 0.041 & Y & Y & Y \\
        \(\beta_{2}\) & 1 & 1.12 & 0.118 & N & Y & Y \\
        \(\beta_{3,bus}\) & 0.6 & 0.638 & 0.063 & Y & Y & Y \\
        \(\beta_{3,metro}\) & 0.3 & 0.314 & 0.048 & Y & Y & Y \\
        \(\beta_{4}\) & 0.6 & 0.643 & 0.072 & Y & Y & Y \\
        \(\gamma_{1,car}\) & -0.1 & -0.091 & 0.092 & Y & Y & Y \\
        \(\gamma_{1,bus}\) & 0.4 & 0.424 & 0.059 & Y & Y & Y \\
        \(\gamma_{1,metro}\) & 0.2 & 0.202 & 0.009 & Y & Y & Y \\
        \(\gamma_{1,bus+metro}\) & 0.25 & 0.251 & 0.006 & Y & Y & Y \\
        \(\gamma_{1,car+bus}\) & 0.2 & 0.198 & 0.011 & Y & Y & Y \\
        \(\gamma_{1,car+metro}\) & 0.15 & 0.154 & 0.028 & Y & Y & N \\
        \(\gamma_{1,car+bus+metro}\) & 0.2 & 0.205 & 0.025 & Y & Y & Y \\
        \(\gamma_{2,car}\) & 0.1 & 0.129 & 0.291 & N & Y & Y \\
        \(\gamma_{2,bus}\) & 0.3 & 0.313 & 0.044 & Y & Y & Y \\
        \(\gamma_{2,metro}\) & 0.2 & 0.202 & 0.010 & Y & Y & Y \\
        \(\gamma_{2,bus+metro}\) & 0.25 & 0.252 & 0.009 & Y & Y & Y \\
        \(\gamma_{2,car+bus}\) & 0.2 & 0.198 & 0.008 & Y & Y & Y \\
        \(\gamma_{2,car+metro}\) & 0.15 & 0.155 & 0.032 & Y & Y & Y \\
        \(\gamma_{2,car+bus+metro}\) & 0.2 & 0.206 & 0.031 & Y & Y & Y \\
        \(\gamma_{3,car}\) & 0.2 & 0.240 & 0.200 & N & Y & Y \\
        \(\gamma_{3,bus}\) & 0 & 0.084 & - & - & - & N \\
        \(\gamma_{3,metro}\) & 0 & 0.082 & - & - & - & N \\
        \(\gamma_{3,bus+metro}\) & 0 & 0.078 & - & - & - & N \\
        \(\gamma_{3,car+bus}\) & 0 & 0.075 & - & - & - & N \\
        \(\gamma_{3,car+metro}\) & 0 & 0.089 & - & - & - & N \\
        \(\gamma_{3,car+bus+metro}\) & 0 & 0.085 & - & - & - & N \\
        \(\alpha_{bus}\) & 4 & 4.324 & 0.081 & Y & Y & Y \\
        \(\alpha_{metro}\) & 2 & 2.111 & 0.055 & Y & Y & Y \\
        \(\alpha_{bus+metro}\) & 2.8 & 2.969 & 0.060 & Y & Y & Y \\
        \(\alpha_{car+bus}\) & 3.2 & 3.416 & 0.067 & Y & Y & Y \\
        \(\alpha_{car+metro}\) & 1.6 & 1.705 & 0.066 & Y & Y & Y \\
        \(\alpha_{car+bus+metro}\) & 3.2 & 3.446 & 0.078 & Y & Y & Y \\
        \hline
        * \footnotesize{Y: yes; N: no} & & & & & & 
    \end{tabular}
\end{table}

We summarize two potential reasons for the inconsistent results of the statistical significance in this experiment. First, the network is on a large scale, but there is only one short metro line in the center of the city. This results in the absence of second-level modes involving metro (especially metro only and car+metro) for many O–D pairs. Specifically, only 0.02\% of the paths for the first-level mode \enquote{public transit} correspond to the second-level mode \enquote{metro only}, and only 0.34\% of the paths for the first-level mode PNR correspond to second-level mode \enquote{car+metro}. This causes a severe imbalance in the mode choice sets and a very low representation of these second-level modes, which can easily lead to statistical inaccuracies in hypothesis testing. Second, due to the high dimensionality and complexity of the problem, the model may need to be downsized. Later in \Cref{sec-Discussions}, we further discuss the impact of model size on hypothesis testing. 

Note that the hypothesis tests conducted on the two networks are not intended to gain real-world insights into travel behavior, as we lack real-world data. Rather, the proposed hypothesis testing framework is intended to demonstrate a general procedure for model and feature selection with statistical rigor, which is conducted simultaneously with the joint estimation of O-D demand and disutility parameters when real-world system-level data are available.

\FloatBarrier

\section{Discussions} \label{sec-Discussions}
This section further explores and summarizes some insights derived from the results of experiments. Meanwhile, to further explore the role of initial O-D demand in the estimation, the non-uniqueness of the solution, and how hypothesis testing works for more condensed disutility functions, we also conduct another experiment on the Nguyen-Dupuis network using different initial O-D demand and a simplified disutility function that considers only the time-related and monetary costs. That is, only parameters in \(\pmb{\beta}\) are being estimated and tested. 

(1) \textit{Information about the ``shape'' of the ground-truth O-D demand is important.}

From \Cref{fig-ND results - OD comparison}, which compares the ground-truth, initial, and estimated O–D demand for the Nguyen–Dupuis network, we observe that the initial O–D demand captures the rough “shape’’ of the ground-truth O–D demand. In other words, it provides the model with some indication of the relative distribution (large or small) of O–D demand across O–D pairs and time intervals, although it does not necessarily reflect the exact magnitude. Such information can guide the model toward the ground-truth O–D demand by providing a better initial point in the solution space.

We also observe larger discrepancies in certain results, such as the 1st, 2nd, and 3rd time intervals of O–D pair 4. This suggests that deviations from the ground truth in a subset of the O–D demand, whether spatially or temporally, do not prevent the model from fitting the observed data well. Essentially, this may indicate the non-uniqueness of the solution, which we further discuss in the following based on an additional set of results.

(2) \textit{Deviating from the ground truth can still result in a high goodness of fit to the observed data.}

\begin{figure}[ht]
    \centering
    \includegraphics[width=0.99\textwidth]{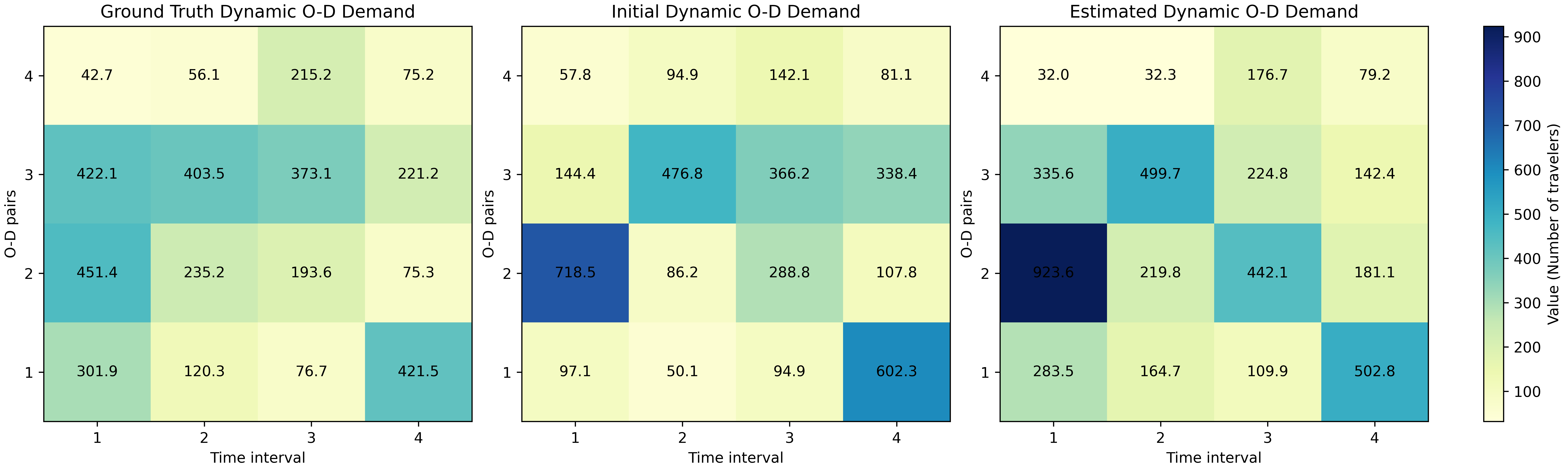}
    \caption{Comparison of ground truth, initial, and estimated dynamic O-D demand (Nguyen-Dupuis, larger noise in initial O-D demand)}
    \label{fig-ND results - OD comparison-small model} 
\end{figure}

\begin{figure}[ht]
    \centering
    \includegraphics[width=1\textwidth]{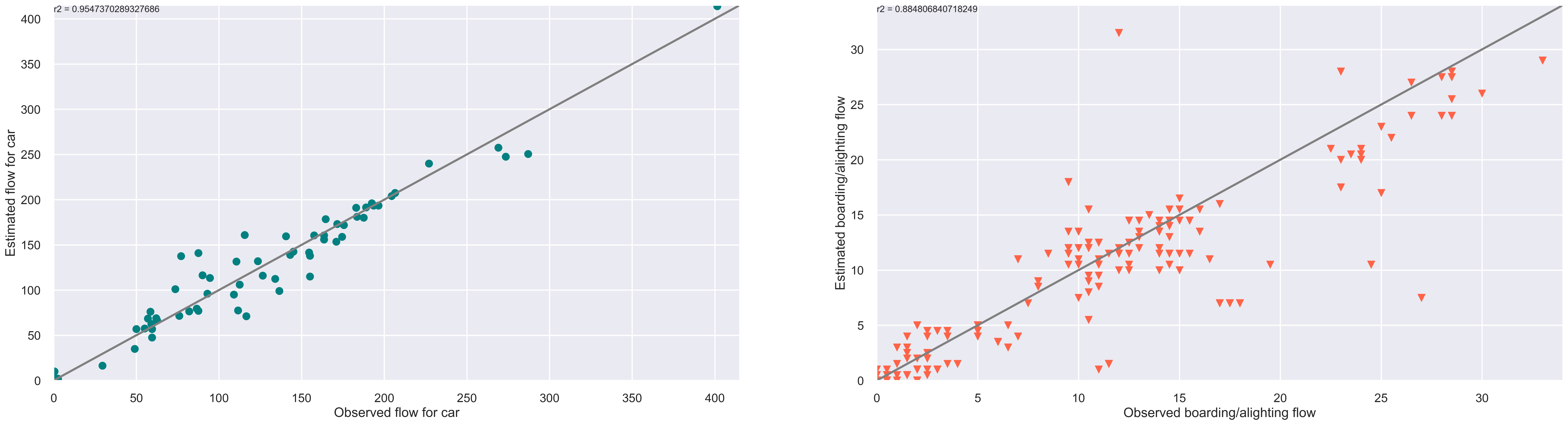}
    \caption{Goodness of fit of link flow (unit: count per 15 minutes) and trip-stop-level bus/metro boarding/alighting counts (Nguyen-Dupuis)}
    \label{fig-ND results - flow fitting - small model} 
\end{figure}

\begin{figure}[ht]
    \centering
    \includegraphics[width=0.68\textwidth]{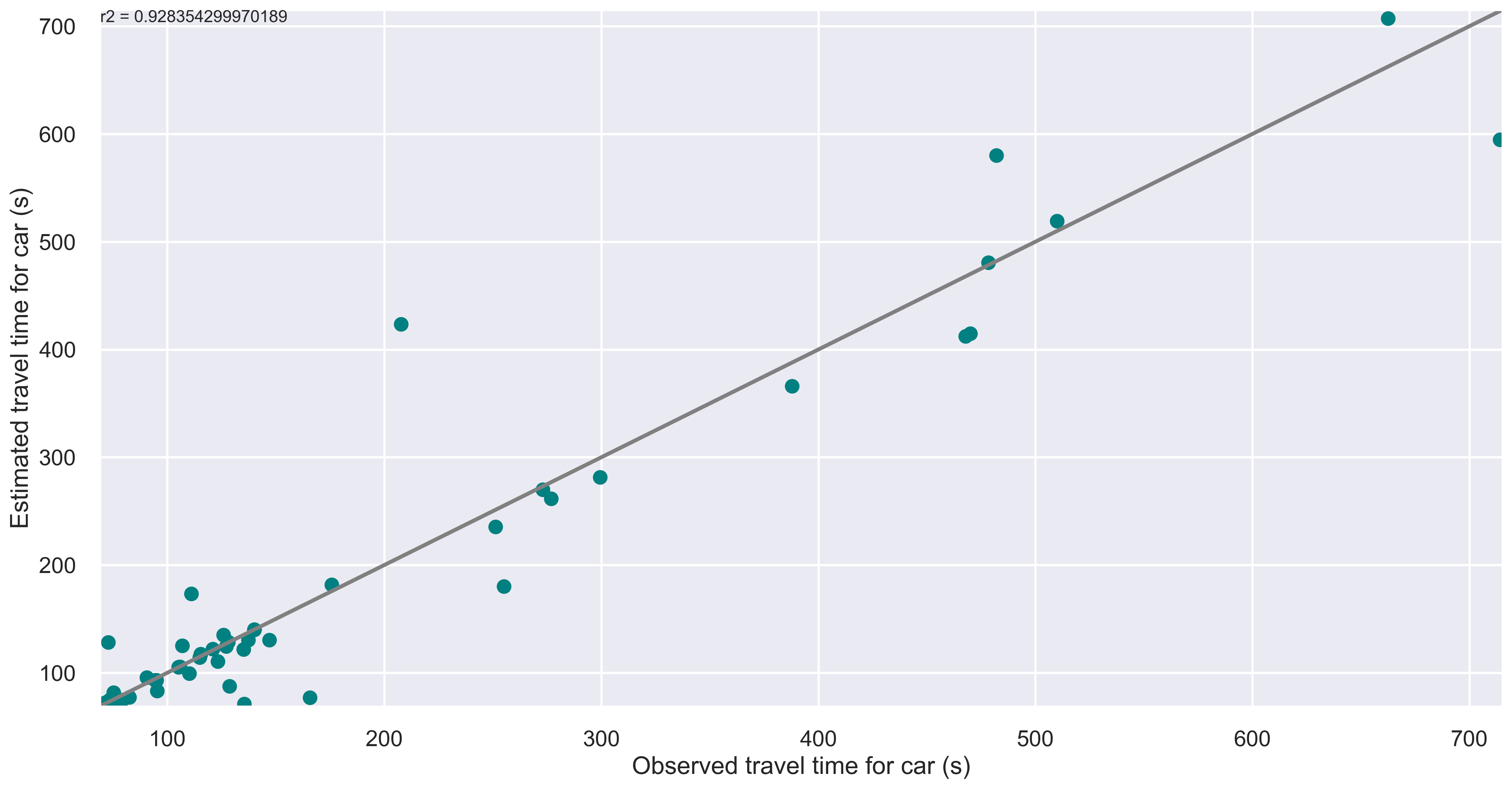}
    \caption{Goodness of fit of dynamic link travel time (Nguyen-Dupuis)}
    \label{fig-ND results - tt fitting - small model} 
\end{figure}

\Cref{fig-ND results - OD comparison-small model} shows the O-D demand estimation results of a different setting on Nguyen-Dupuis network as mentioned earlier. In this case, the random noise used to generate the initial O-D demand is larger, i,e., [-70\%, 70\%]. \Cref{fig-ND results - flow fitting - small model} and \Cref{fig-ND results - tt fitting - small model} present the goodness-of-fit results for the three types of observed data. 


From these results, we can see that (1) when a larger noise is added in the initial O-D demand, the ``shape'' information is no longer consistent with the ground-truth, which provides little value in guiding the learning towards the ground truth; (2) when the estimated O-D demand deviates substantially from the ground truth, e.g., some learned value is nearly twice the ground truth (1st time interval of O-D pair 2), the observed data can still be matched well (see \Cref{fig-ND results - flow fitting - small model} and \Cref{fig-ND results - tt fitting - small model}). This implies that the estimated dynamic O–D demand does not necessarily need to be close to the ground truth to fit the data. Similar findings were observed earlier in the estimation of disutility parameters. 

From a practical perspective, this means that we can observe multiple non-unique sets of O-D demand and parameters that lead to satisfactory performance on data matching. And these non-unique solutions (could be global or local optimal mathematically) can deviate from each other to a potentially substantial extent. 

These implications raise several thought-provoking questions: (a) Does simply achieving a high goodness of fit provide sufficient confidence for subsequent tasks such as what-if analyses for policy insights? (b) Are the exact values of the estimated O–D demand and parameters important, or can they be treated as latent variables representing an “unknown world’’ for analysis purposes? Further research and experimentation are needed to answer these questions. However, we can at least conclude that obtaining estimates closer to the ground truth requires an initial starting point that contains rough “shape’’ information. Therefore, given the complexity of the problem and the non-uniqueness of the solutions, it is recommended to leverage as much prior information as possible when performing the estimation. In practice, such information can be obtained from various data sources, including planning agencies and data vendors.

(3) \textit{Matching the trip-and-stop-level passenger boarding/alighting counts is more challenging than matching the car flow.}

From the results of \Cref{fig-ND results - flow fitting}, \Cref{fig-Genoa results - flow fitting}, and \Cref{fig-ND results - flow fitting - small model}, we can see that fitting the trip-and-stop-level passenger boarding/alighting counts is more challenging than fitting the car flow. This is because: 
\begin{enumerate}[label=\alph*),leftmargin=3.8em, itemsep=0.2em]
    \item The temporal and spatial space is more discrete when fitting passenger counts, and the optimization operates at a more granular level. The optimization requires matching at the trip-and-stop level. Bus and metro vehicles arrive in a discrete temporal space, where small discrepancies in travel time can lead to large inconsistencies in determining which specific trip (i.e., the specific vehicle run for the same bus or metro line) passengers are modeled to take, thereby affecting their perceived costs and subsequent mode/route choices.
    \item It is more challenging to obtain a highly accurate DAR matrix for passengers when computing the related derivatives in the computational graph. Discretization is required to represent the dynamic path flows (the columns of the DAR matrix). However, it is very difficult to accurately determine the percentage of passengers using a particular path who board or alight at a specific stop and on a specific bus or metro vehicle. The DAR is sensitive to how passenger demand is distributed within a single time interval. Reducing the length of the discretized time interval could improve accuracy, but doing so introduces substantial computational burden, simulation instability, and a higher likelihood of numerical errors.
    \item The results are sensitive to multiple factors, such as the accuracy of travel time estimation and the passenger path table used, when matching trip-and-stop-level passenger boarding/alighting counts. As mentioned earlier, small discrepancies in travel time can lead to large inconsistencies in the specific trip that passengers are modeled to take (e.g., missing a bus/metro vehicle and boarding a different one). Additionally, inaccuracies in the passenger path table can easily create conflicts when attempting to match the observed data. For example, suppose in the path table, stop A and stop B are highly and positively correlated (i.e., these two stops are typically used together, so increasing the ridership at one stop automatically increases the ridership at the other). In reality, however, many other factors such as alternative path choices and random unobserved behaviors may weaken or eliminate such a correlation. This mismatch can create conflicts and impede the ability to match observed data. More discussion on the passenger path table is presented in point (5) below.
\end{enumerate}

(4) \textit{Hypothesis testing works well on condensed disutility functions for non-linear and complex estimation problems.}

Hypothesis testing has been shown to work well for simple linear models, such as ordinary least squares (OLS). However, its capability in large-scale, high-dimensional, nonlinear problems (or even more ill-conditioned problems) remains unclear. In transportation research, there is a strong need for tools that can validate whether learned parameters are statistically significant, especially in complex estimation settings. This research makes such an attempt by proposing a hypothesis testing framework for a joint estimation problem that is highly complex, involving both continuous and discrete components across temporal and spatial dimensions.

The results in the previous section were obtained using a full disutility function containing a large number of factors and parameters, which will likely need to be condensed when working with real-world data (e.g., due to correlations among different factors, data availability issues, or data biases). Although the earlier results demonstrated certain capability of the framework under the full model, here we further examine whether the statistical significance results can become more consistent with the assumed ground truth when a simplified disutility function is adopted for this complex estimation problem. \Cref{tab - hypo ND - small model} shows such consistency, indicating the capability and promise of the proposed hypothesis testing framework. For real-world tasks, if using a full model of disutility function with a large number of factors, multiple times of iterative estimation and testing might be also helpful. 

\begin{table}[!htbp]
    \centering
    \caption{Disutility parameter estimation and hypothesis testing results (Nguyen-Dupuis, simplified disutility function)}
    \label{tab - hypo ND - small model}  
    \begin{tabular}{l l l l l l l l}
        \hline
        Parameter & \begin{tabular}{@{}l@{}}Ground truth \\ value\end{tabular} & \begin{tabular}{@{}l@{}}Estimated \\ value\end{tabular} & \begin{tabular}{@{}l@{}l@{}}Absolute \\ percentage \\ error \end{tabular} & \begin{tabular}{@{}l@{}}Right \\ direction\end{tabular} & \begin{tabular}{@{}l@{}}Right \\ sign\end{tabular} & \begin{tabular}{@{}l@{}}Consistent \\ significance\end{tabular}\\
        \hline
        \(\beta_{1,car}\) & 1 & 1.144 & 0.144 & N & Y & Y \\
        \(\beta_{1,bus}\) & 1.5 & 1.624 & 0.083 & Y & Y & Y \\
        \(\beta_{1,metro}\) & 0.8 & 0.889 & 0.111 & N & Y & Y \\
        \(\beta_{2}\) & 2 & 2.218 & 0.109 & N & Y & Y \\
        \(\beta_{3,bus}\) & 2 & 2.2 & 0.100 & - & Y & Y \\
        \(\beta_{3,metro}\) & 1 & 1.124 & 0.124 & N & Y & Y \\
        \(\beta_{4}\) & 2 & 2.153 & 0.077 & Y & Y & Y \\
        \hline
        * \footnotesize{Y: yes; N: no} & & & & & & 
    \end{tabular}
\end{table}

(5) \textit{Meticulous generation of the passenger path table is important for real-world tasks.}

One limitation of the present research is that we use a pre-defined path table for modeling public transit passengers' behaviors. This may not be a critical issue when using synthetic data since as long as the synthetic data is generated using the same simulation tool we can avoid the trouble that should be caused by any inconsistencies in the path table compared to the ground truth. However, when it comes to fitting the observed data from the real world, careful attention should be paid into the path table generation.

We propose three potential solutions: (1) Leverage heuristics to dynamically generate the path table across iterations in the estimation process, enabling broader coverage of feasible passenger paths based on real-time travel costs, for example, through column generation; (2) Use well-generated path tables from navigation applications, as these are generally well aligned with reality and reflect real-time dynamics of public transit schedules. Such path tables also tend to incorporate a reasonable level of heterogeneity within the same O–D pair, which is often overlooked when using simple shortest-path algorithms; (3) Adopt dynamic path tables for each time interval, allowing the model to capture a time-varying set of paths that are “optimal’’ both spatially and temporally.

\section{Conclusions} \label{sec: conclusions}
This research proposes a joint estimation framework for dynamic origin–destination (O–D) demand and disutility functions within a multi-modal transportation system, along with a hypothesis testing framework for evaluating the statistical significance of the estimated parameters. The framework leverages multi-source system-level data, including traffic counts, speeds, and trip-and-stop-level public transit boarding and alighting data, within the estimation process. A generalized and comprehensive disutility specification, incorporating both alternative-specific and zone-specific factors, is embedded into hierarchical disutility functions to capture heterogeneous traveler perceptions. The framework integrates a multi-modal dynamic traffic assignment model that simultaneously captures route and mode choices and features detailed, mode-specific travel time modeling. The estimation problem is formulated and solved using a computational graph-based learning approach, enabling dynamic network modeling and scalable inference on large-scale networks. Furthermore, we develop a hypothesis testing framework tailored to this complex estimation setting, allowing the statistical significance of behavioral parameters to be assessed, supporting model and feature selection, and informing policy implications for real-world applications. Experiments on both a toy network and a large-scale real-world network validate the effectiveness of the proposed framework and demonstrate satisfactory performance. Additional analyses and discussions provide rich insights into real-world estimation challenges and highlight several important research questions. Overall, this work contributes a scalable, data-driven methodology for multi-modal demand and behavioral parameter estimation and offers the possibility for statistically rigorous insights for decision-makers.

There are several directions for future research. First, evaluating the impact of passenger path table generation on estimation performance, particularly when using real-world data, remains an important topic. Second, the interplay between O–D demand estimation and the estimation of disutility parameters warrants further investigation, as understanding how these two components influence each other could yield valuable practical insights. Third, computing gradients in this framework introduces high complexity; thus, future work could explore training neural networks to represent traffic dynamics, thereby taking full advantage of auto-differentiation and modern deep learning tools.

\section*{Acknowledgments}
This research is sponsored by Hitachi Rail. The authors thank Hitachi Rail for providing the relevant data of the Genoa network. The contents of this article
reflect the views of the authors only, who are responsible for the facts and accuracy of the information presented herein. 

\section*{Appendix A}
\subsection*{(1) Derivation of \( \frac{\partial f^{rs}_{m,g(m),k,t}}{\partial c^{rs}_{m,g(m),k,t}} \)}

This case considers the derivative of path flow with respect to the disutility of the same path \(k\).

Since \( f^{rs}_{m,g(m),k,t} = q^{rs}_{t} \cdot \mathcal{P}^{rs}_{m,t} \cdot \mathcal{P}^{rs}_{g(m)|m,t} \cdot \mathcal{P}^{rs}_{k|g(m),t} \), we have
\begin{align}
\frac{\partial f^{rs}_{m,g(m),k,t}}{\partial c^{rs}_{m,g(m),k,t}} &= q^{rs}_{t} \cdot \Big(
\frac{\partial \mathcal{P}^{rs}_{m,t}}{\partial c^{rs}_{m,g(m),k,t}} \cdot \mathcal{P}^{rs}_{g(m)|m,t} \cdot \mathcal{P}^{rs}_{k|g(m),t} + \mathcal{P}^{rs}_{m,t} \cdot \frac{\partial \mathcal{P}^{rs}_{g(m)|m,t}}{\partial c^{rs}_{m,g(m),k,t}} \cdot \mathcal{P}^{rs}_{k|g(m),t} \nonumber \\
&\quad+ \mathcal{P}^{rs}_{m,t} \cdot \mathcal{P}^{rs}_{g(m)|m,t} \cdot \frac{\partial \mathcal{P}^{rs}_{k|g(m),t}}{\partial c^{rs}_{m,g(m),k,t}} \Big)
\label{eq: der of f wrt c}
\end{align}

\paragraph{(i) Derivative of \( \mathcal{P}^{rs}_{k|g(m),t} \)}
\begin{align}
\frac{\partial \mathcal{P}^{rs}_{k|g(m),t}}{\partial c^{rs}_{m,g(m),k,t}} &= \frac{\partial \mathcal{P}^{rs}_{k|g(m),t}}{\partial e^{-\frac{1}{\theta_{m,g(m)}}c^{rs}_{m,g(m),k,t}}} \cdot \frac{\partial e^{-\frac{1}{\theta_{m,g(m)}}c^{rs}_{m,g(m),k,t}}}{\partial c^{rs}_{m,g(m),k,t}} \nonumber\\
& = \frac{\sum_{k \in P^{rs}_{m,g(m)}}e^{-\frac{1}{\theta_{m,g(m)}}c^{rs}_{m,g(m),k,t}} - e^{-\frac{1}{\theta_{m,g(m)}}c^{rs}_{m,g(m),k,t}}}{\left(\sum_{k \in P^{rs}_{m,g(m)}}e^{-\frac{1}{\theta_{m,g(m)}}c^{rs}_{m,g(m),k,t}}\right)^2} \cdot \left(-\frac{1}{\theta_{m,g(m)}}e^{-\frac{1}{\theta_{m,g(m)}}c^{rs}_{m,g(m),k,t}}\right)\nonumber\\
& = -\frac{1}{\theta_{m,g(m)}} \cdot \frac{\sum_{k \in P^{rs}_{m,g(m)}}e^{-\frac{1}{\theta_{m,g(m)}}c^{rs}_{m,g(m),k,t}} - e^{-\frac{1}{\theta_{m,g(m)}}c^{rs}_{m,g(m),k,t}}}{\sum_{k \in P^{rs}_{m,g(m)}}e^{-\frac{1}{\theta_{m,g(m)}}c^{rs}_{m,g(m),k,t}}} \cdot \frac{e^{-\frac{1}{\theta_{m,g(m)}}c^{rs}_{m,g(m),k,t}}}{\sum_{k \in P^{rs}_{m,g(m)}}e^{-\frac{1}{\theta_{m,g(m)}}c^{rs}_{m,g(m),k,t}}}\nonumber\\
& = -\frac{1}{\theta_{m,g(m)}} \mathcal{P}^{rs}_{k|g(m),t}(1 - \mathcal{P}^{rs}_{k|g(m),t})
\label{eq: der of P_k}
\end{align}

\paragraph{(ii) Derivative of \( \mathcal{P}^{rs}_{g(m)|m,t} \)}
\begin{align}
\frac{\partial \mathcal{P}^{rs}_{g(m)|m,t}}{\partial c^{rs}_{m,g(m),k,t}} &= \frac{\partial \mathcal{P}^{rs}_{g(m)|m,t}}{\partial IV^{rs}_{m,g(m),t}} \cdot \frac{\partial IV^{rs}_{m,g(m),t}}{\partial c^{rs}_{m,g(m),k,t}} \nonumber\\
& = \frac{\theta_{m,g(m)}}{\theta_m} \mathcal{P}^{rs}_{g(m)|m,t}(1 - \mathcal{P}^{rs}_{g(m)|m,t}) \cdot \left(\frac{e^{-\frac{1}{\theta_{m,g(m)}}c^{rs}_{m,g(m),k,t}}}{\sum_{k \in P^{rs}_{m,g(m)}}e^{-\frac{1}{\theta_{m,g(m)}}c^{rs}_{m,g(m),k,t}}} \cdot \left(-\frac{1}{\theta_{m,g(m)}}\right) \right) \nonumber \\
& = \frac{\theta_{m,g(m)}}{\theta_m} \mathcal{P}^{rs}_{g(m)|m,t}(1 - \mathcal{P}^{rs}_{g(m)|m,t}) \cdot \left( -\frac{1}{\theta_{m,g(m)}} \mathcal{P}^{rs}_{k|g(m),t} \right) \nonumber \\
&= -\frac{1}{\theta_m} \mathcal{P}^{rs}_{g(m)|m,t}(1 - \mathcal{P}^{rs}_{g(m)|m,t}) \mathcal{P}^{rs}_{k|g(m),t}
\label{eq: der of P_g(m)}
\end{align}

\paragraph{(iii) Derivative of \( \mathcal{P}^{rs}_{m,t} \)}
\begin{align}
\frac{\partial \mathcal{P}^{rs}_{m,t}}{\partial c^{rs}_{m,g(m),k,t}} &= \frac{\partial \mathcal{P}^{rs}_{m,t}}{\partial IV^{rs}_{m,t}} \cdot \frac{\partial IV^{rs}_{m,t}}{\partial IV^{rs}_{m,g(m),t}} \cdot \frac{\partial IV^{rs}_{m,g(m),t}}{\partial c^{rs}_{m,g(m),k,t}} \nonumber\\
& = \theta_m \mathcal{P}^{rs}_{m,t}(1 - \mathcal{P}^{rs}_{m,t}) \cdot \frac{\theta_{m,g(m)}}{\theta_m}\mathcal{P}^{rs}_{g(m)|m,t} \cdot \left(-\frac{1}{\theta_{m,g(m)}}\right)\mathcal{P}^{rs}_{k|g(m),t} \nonumber \\
& = -\mathcal{P}^{rs}_{m,t}(1 - \mathcal{P}^{rs}_{m,t}) \mathcal{P}^{rs}_{g(m)|m,t} \mathcal{P}^{rs}_{k|g(m),t}
\label{eq: der of P_m}
\end{align}

Therefore, according to Eqs. \eqref{eq: der of f wrt c} - \eqref{eq: der of P_m}, we have
\begin{align*}
\frac{\partial f^{rs}_{m,g(m),k,t}}{\partial c^{rs}_{m,g(m),k,t}} &= f^{rs}_{m,g(m),k,t} \cdot \Bigg[
-\frac{1}{\theta_{m,g(m)}} (1 - \mathcal{P}^{rs}_{k|g(m),t})
- \frac{1}{\theta_m} \mathcal{P}^{rs}_{k|g(m),t}(1 - \mathcal{P}^{rs}_{g(m)|m,t}) \\
&\quad - \mathcal{P}^{rs}_{k|g(m),t} \mathcal{P}^{rs}_{g(m)|m,t} (1 - \mathcal{P}^{rs}_{m,t})
\Bigg]
\end{align*}

\subsection*{(2) Derivation of \( \frac{\partial f^{rs}_{m,g(m),k,t}}{\partial c^{rs}_{m,g(m),k',t}} \) for \( k' \neq k \)}

This case considers the derivative with respect to the disutility of a different path \(k'\) within the same second-level mode \(g(m)\). We have
\begin{align}
\frac{\partial f^{rs}_{m,g(m),k,t}}{\partial c^{rs}_{m,g(m),k',t}} &= q^{rs}_{t} \cdot \Big(
\frac{\partial \mathcal{P}^{rs}_{m,t}}{\partial c^{rs}_{m,g(m),k',t}} \cdot \mathcal{P}^{rs}_{g(m)|m,t} \cdot \mathcal{P}^{rs}_{k|g(m),t} + \mathcal{P}^{rs}_{m,t} \cdot \frac{\partial \mathcal{P}^{rs}_{g(m)|m,t}}{\partial c^{rs}_{m,g(m),k',t}} \cdot \mathcal{P}^{rs}_{k|g(m),t} \nonumber \\
&\quad+ \mathcal{P}^{rs}_{m,t} \cdot \mathcal{P}^{rs}_{g(m)|m,t} \cdot \frac{\partial \mathcal{P}^{rs}_{k|g(m),t}}{\partial c^{rs}_{m,g(m),k',t}} \Big)
\label{eq: case 2: der of f wrt c_k'}
\end{align}

\paragraph{(i) Derivative of \( \mathcal{P}^{rs}_{k|g(m),t} \)}

\begin{align}
\frac{\partial \mathcal{P}^{rs}_{k|g(m),t}}{\partial c^{rs}_{m,g(m),k',t}} 
& = e^{-\frac{1}{\theta_{m,g(m)}}c^{rs}_{m,g(m),k,t}} \cdot (-1) \left(\sum_{k \in P^{rs}_{m,g(m)}}e^{-\frac{1}{\theta_{m,g(m)}}c^{rs}_{m,g(m),k,t}}\right)^{-2} \cdot \left(-\frac{1}{\theta_{m,g(m)}}e^{-\frac{1}{\theta_{m,g(m)}}c^{rs}_{m,g(m),k',t}}\right)\nonumber\\
& = \frac{1}{\theta_{m,g(m)}} \cdot \frac{e^{-\frac{1}{\theta_{m,g(m)}}c^{rs}_{m,g(m),k,t}}}{\sum_{k \in P^{rs}_{m,g(m)}}e^{-\frac{1}{\theta_{m,g(m)}}c^{rs}_{m,g(m),k,t}}} \cdot \frac{e^{-\frac{1}{\theta_{m,g(m)}}c^{rs}_{m,g(m),k',t}}}{\sum_{k \in P^{rs}_{m,g(m)}}e^{-\frac{1}{\theta_{m,g(m)}}c^{rs}_{m,g(m),k,t}}}\nonumber\\
& = \frac{1}{\theta_{m,g(m)}} \mathcal{P}^{rs}_{k|g(m),t}\mathcal{P}^{rs}_{k'|g(m),t}
\label{eq: case 2: der of P_k wrt c_k'}
\end{align}

\paragraph{(ii) Derivative of \( \mathcal{P}^{rs}_{g(m)|m,t} \)}

\begin{align}
\frac{\partial \mathcal{P}^{rs}_{g(m)|m,t}}{\partial c^{rs}_{m,g(m),k',t}} &= \frac{\partial \mathcal{P}^{rs}_{g(m)|m,t}}{\partial IV^{rs}_{m,g(m),t}} \cdot \frac{\partial IV^{rs}_{m,g(m),t}}{\partial c^{rs}_{m,g(m),k',t}} \nonumber\\
& = \frac{\theta_{m,g(m)}}{\theta_m} \mathcal{P}^{rs}_{g(m)|m,t}(1 - \mathcal{P}^{rs}_{g(m)|m,t}) \cdot \left( -\frac{1}{\theta_{m,g(m)}} \mathcal{P}^{rs}_{k'|g(m),t} \right) \nonumber \\
&= -\frac{1}{\theta_m} \mathcal{P}^{rs}_{g(m)|m,t}(1 - \mathcal{P}^{rs}_{g(m)|m,t}) \mathcal{P}^{rs}_{k'|g(m),t}
\label{eq: case 2: der of P_g(m) wrt c_k'}
\end{align}

\paragraph{(iii) Derivative of \( \mathcal{P}^{rs}_{m,t} \)}

\begin{align}
\frac{\partial \mathcal{P}^{rs}_{m,t}}{\partial c^{rs}_{m,g(m),k',t}} &= \frac{\partial \mathcal{P}^{rs}_{m,t}}{\partial IV^{rs}_{m,t}} \cdot \frac{\partial IV^{rs}_{m,t}}{\partial IV^{rs}_{m,g(m),t}} \cdot \frac{\partial IV^{rs}_{m,g(m),t}}{\partial c^{rs}_{m,g(m),k',t}} \nonumber\\
& = \theta_m \mathcal{P}^{rs}_{m,t}(1 - \mathcal{P}^{rs}_{m,t}) \cdot \frac{\theta_{m,g(m)}}{\theta_m}\mathcal{P}^{rs}_{g(m)|m,t} \cdot \left(-\frac{1}{\theta_{m,g(m)}}\right)\mathcal{P}^{rs}_{k'|g(m),t} \nonumber \\
& = -\mathcal{P}^{rs}_{m,t}(1 - \mathcal{P}^{rs}_{m,t}) \mathcal{P}^{rs}_{g(m)|m,t} \mathcal{P}^{rs}_{k'|g(m),t}
\label{eq: case 2: der of P_m wrt c_k'}
\end{align}

Therefore, according to Eqs. \eqref{eq: case 2: der of f wrt c_k'} - \eqref{eq: case 2: der of P_m wrt c_k'}, we have
\begin{align*}
\frac{\partial f^{rs}_{m,g(m),k,t}}{\partial c^{rs}_{m,g(m),k',t}} &= f^{rs}_{m,g(m),k,t} \cdot \Bigg[
\frac{1}{\theta_{m,g(m)}} \mathcal{P}^{rs}_{k'|g(m),t}
- \frac{1}{\theta_m} \mathcal{P}^{rs}_{k'|g(m),t}(1 - \mathcal{P}^{rs}_{g(m)|m,t}) \\
&\quad - \mathcal{P}^{rs}_{k'|g(m),t} \mathcal{P}^{rs}_{g(m)|m,t} (1 - \mathcal{P}^{rs}_{m,t})
\Bigg]
\end{align*}

\subsection*{(3) Derivation of \( \frac{\partial f^{rs}_{m,g(m),k,t}}{\partial c^{rs}_{m,g'(m),k',t}} \) for \( k \neq k', g'(m) \neq g(m) \)}

This case considers the derivative with respect to the disutility of path \(k'\) in a different second-level mode \(g'(m)\) but within the same first-level mode \(m\).

\paragraph{(i) Derivative of \( \mathcal{P}^{rs}_{k|g(m),t} \)} \mbox{}\\

The disutility \(c^{rs}_{m,g'(m),k',t}\) is not in the choice set of nest \(g(m)\). Therefore,
\begin{equation}
\frac{\partial \mathcal{P}^{rs}_{k|g(m),t}}{\partial c^{rs}_{m,g'(m),k',t}} = 0
\label{eq: case 3: der of P_k wrt c_k'}
\end{equation}

\paragraph{(ii) Derivative of \( \mathcal{P}^{rs}_{g(m)|m,t} \)}

\begin{align}
\frac{\partial \mathcal{P}^{rs}_{g(m)|m,t}}{\partial c^{rs}_{m,g'(m),k',t}} &= - \mathcal{P}^{rs}_{g(m)|m,t} \mathcal{P}^{rs}_{g'(m)|m,t} \cdot \frac{\partial (\frac{\theta_{m,g'(m)}}{\theta_m}IV^{rs}_{m,g'(m),t})}{\partial c^{rs}_{m,g'(m),k',t}} \nonumber\\
&= - \mathcal{P}^{rs}_{g(m)|m,t} \mathcal{P}^{rs}_{g'(m)|m,t} \cdot \frac{\theta_{m,g'(m)}}{\theta_m} \cdot \frac{\partial IV^{rs}_{m,g'(m),t}}{\partial c^{rs}_{m,g'(m),k',t}} \nonumber\\
&= - \mathcal{P}^{rs}_{g(m)|m,t} \mathcal{P}^{rs}_{g'(m)|m,t} \cdot \frac{\theta_{m,g'(m)}}{\theta_m} \cdot \left( -\frac{1}{\theta_{m,g'(m)}} \mathcal{P}^{rs}_{k'|g'(m),t} \right) \nonumber\\
&= \frac{1}{\theta_m} \mathcal{P}^{rs}_{g(m)|m,t} \mathcal{P}^{rs}_{g'(m)|m,t} \mathcal{P}^{rs}_{k'|g'(m),t}
\label{eq: case 3: der of P_g(m) wrt c_k'}
\end{align}

\paragraph{(iii) Derivative of \( \mathcal{P}^{rs}_{m,t} \)}
\begin{align}
\frac{\partial \mathcal{P}^{rs}_{m,t}}{\partial c^{rs}_{m,g'(m),k',t}} &= \frac{\partial \mathcal{P}^{rs}_{m,t}}{\partial IV^{rs}_{m,t}} \cdot \frac{\partial IV^{rs}_{m,t}}{\partial IV^{rs}_{m,g'(m),t}} \cdot \frac{\partial IV^{rs}_{m,g'(m),t}}{\partial c^{rs}_{m,g'(m),k',t}} \nonumber\\
&= \theta_m \mathcal{P}^{rs}_{m,t}(1 - \mathcal{P}^{rs}_{m,t}) \cdot \frac{\theta_{m,g'(m)}}{\theta_m}\mathcal{P}^{rs}_{g'(m)|m,t} \cdot \left(-\frac{1}{\theta_{m,g'(m)}}\right)\mathcal{P}^{rs}_{k'|g'(m),t} \nonumber \\
&= -\mathcal{P}^{rs}_{m,t}(1 - \mathcal{P}^{rs}_{m,t}) \mathcal{P}^{rs}_{g'(m)|m,t} \mathcal{P}^{rs}_{k'|g'(m),t}
\label{eq: case 3: der of P_m wrt c_k'}
\end{align}

Therefore, according to Eqs. \eqref{eq: case 3: der of P_k wrt c_k'} - \eqref{eq: case 3: der of P_m wrt c_k'}, we have
\begin{align*}
\frac{\partial f^{rs}_{m,g(m),k,t}}{\partial c^{rs}_{m,g'(m),k',t}} &= f^{rs}_{m,g(m),k,t} \cdot \mathcal{P}^{rs}_{g'(m)|m,t} \cdot \mathcal{P}^{rs}_{k'|g'(m),t} \cdot \Bigg(
\frac{1}{\theta_m}  - 1 + \mathcal{P}^{rs}_{m,t}\Bigg)
\end{align*}

\subsection*{(4) Derivation of \( \frac{\partial f^{rs}_{m,g(m),k,t}}{\partial c^{rs}_{m',g'(m'),k',t}} \) for \(k \neq k', g'(m') \neq g(m), m' \neq m \)}

This case considers the derivative with respect to path \(k'\) in an entirely different first-level mode \(m'\).

\paragraph{(i) Derivative of \( \mathcal{P}^{rs}_{k|g(m),t} \)}
\begin{equation}
\frac{\partial \mathcal{P}^{rs}_{k|g(m),t}}{\partial c^{rs}_{m',g'(m'),k',t}} = 0
\label{eq: case 4: der of P_k wrt c_k'}
\end{equation}

\paragraph{(ii) Derivative of \( \mathcal{P}^{rs}_{g(m)|m,t} \)}
\begin{equation}
\frac{\partial \mathcal{P}^{rs}_{g(m)|m,t}}{\partial c^{rs}_{m',g'(m'),k',t}} = 0
\label{eq: case 4: der of P_g(m) wrt c_k'}
\end{equation}

\paragraph{(iii) Derivative of \( \mathcal{P}^{rs}_{m,t} \)}
\begin{align}
\frac{\partial \mathcal{P}^{rs}_{m,t}}{\partial c^{rs}_{m',g'(m'),k',t}} &= - \mathcal{P}^{rs}_{m,t}\mathcal{P}^{rs}_{m',t} \cdot \frac{\partial (\theta_{m'}IV^{rs}_{m',t})}{\partial c^{rs}_{m',g'(m'),k',t}} \nonumber\\
&= - \mathcal{P}^{rs}_{m,t}\mathcal{P}^{rs}_{m',t} \cdot \theta_{m'} \left( \frac{\partial IV^{rs}_{m',t}}{\partial IV^{rs}_{m',g'(m'),t}} \cdot \frac{\partial IV^{rs}_{m',g'(m'),t}}{\partial c^{rs}_{m',g'(m'),k',t}} \right) \nonumber\\
&= - \mathcal{P}^{rs}_{m,t}\mathcal{P}^{rs}_{m',t} \cdot \theta_{m'} \left( \frac{\theta_{m',g'(m')}}{\theta_{m'}} \mathcal{P}^{rs}_{g'(m')|m',t} \cdot \left(-\frac{1}{\theta_{m',g'(m')}}\right) \mathcal{P}^{rs}_{k'|g'(m'),t} \right) \nonumber\\
&= - \mathcal{P}^{rs}_{m,t}\mathcal{P}^{rs}_{m',t} \cdot \left( - \mathcal{P}^{rs}_{g'(m')|m',t}\mathcal{P}^{rs}_{k'|g'(m'),t} \right) \nonumber\\
&= \mathcal{P}^{rs}_{m,t}\mathcal{P}^{rs}_{m',t}\mathcal{P}^{rs}_{g'(m')|m',t}\mathcal{P}^{rs}_{k'|g'(m'),t}
\label{eq: case 4: der of P_m wrt c_k'}
\end{align}

Therefore, according to Eqs. \eqref{eq: case 4: der of P_k wrt c_k'} - \eqref{eq: case 4: der of P_m wrt c_k'}, we have
\begin{align*}
\frac{\partial f^{rs}_{m,g(m),k,t}}{\partial c^{rs}_{m',g'(m'),k',t}} &= f^{rs}_{m,g(m),k,t} \cdot \mathcal{P}^{rs}_{m',t} \cdot \mathcal{P}^{rs}_{g'(m')|m',t} \cdot \mathcal{P}^{rs}_{k'|g'(m'),t}
\end{align*}

\section*{Appendix B}
Nguyen-Dupuis network specifics used in the experiments, including the link characteristics and other information such as parking fees and the social demographic features are shown in \Cref{tab - ND Link characteristics} and \Cref{tab - ND info}.

\begin{table}[H]
    \centering
    \caption{Link characteristics of Nguyen-Dupuis network}
    \label{tab - ND Link characteristics}  
    \begin{tabular}{l l l l l l l l}
        \hline
        Link ID & \begin{tabular}{@{}l@{}}Length\\(mile)\end{tabular} & \begin{tabular}{@{}l@{}l@{}}Speed\\limit \\ (mph)\end{tabular} & \begin{tabular}{@{}l@{}}Capacity\\(car,vph)\end{tabular}& \begin{tabular}{@{}l@{}l@{}}Jam\\density \\ (car/mile)\end{tabular}& \begin{tabular}{@{}l@{}}Capacity\\(bus,vph)\end{tabular} & \begin{tabular}{@{}l@{}l@{}}Jam\\density \\ (bus/mile)\end{tabular} & \begin{tabular}{@{}l@{}}Number of\\lanes\end{tabular} \\
        \hline
        1 & 0.6 & 30 & 900 & 250 & 700 & 200 & 1 \\
        2 & 0.6 & 30 & 900 & 250 & 700 & 200 & 1 \\
        3 & 0.6 & 30 & 900 & 250 & 700 & 200 & 1 \\
        4 & 1.2 & 40 & 1200 & 250 & 1000 & 200 & 2 \\
        5 & 0.6 & 30 & 900 & 250 & 700 & 200 & 1 \\
        6 & 0.6 & 30 & 900 & 250 & 700 & 200 & 1 \\
        7 & 0.6 & 30 & 900 & 250 & 700 & 200 & 1 \\
        8 & 1.2 & 40 & 1200 & 250 & 1000 & 200 & 2 \\
        9 & 0.6 & 30 & 900 & 250 & 700 & 200 & 1 \\
        10 & 2 & 50 & 1600 & 250 & 1400 & 200 & 2 \\
        11 & 0.6 & 30 & 900 & 250 & 700 & 200 & 1 \\
        12 & 0.6 & 30 & 900 & 250 & 700 & 200 & 1 \\
        13 & 0.6 & 30 & 900 & 250 & 700 & 200 & 1 \\
        14 & 0.6 & 30 & 900 & 250 & 700 & 200 & 1 \\
        15 & 0.6 & 30 & 900 & 250 & 700 & 200 & 1 \\
        16 & 0.6 & 30 & 900 & 250 & 700 & 200 & 1 \\
        17 & 0.6 & 30 & 900 & 250 & 700 & 200 & 1 \\
        18 & 0.6 & 30 & 900 & 250 & 700 & 200 & 1 \\
        19 & 0.6 & 30 & 900 & 250 & 700 & 200 & 1 \\
        \hline
    \end{tabular}
\end{table}

\begin{table}[H]
    \centering
    \caption{Other information of Nguyen-Dupuis network}
    \label{tab - ND info}  
    \begin{tabular}{l l l l l}
        \hline
        O/D node ID & \begin{tabular}{@{}l@{}}Median\\income (\$, hourly)\end{tabular} & \begin{tabular}{@{}l@{}l@{}}Population\\density \\ (per \(100m^2\))\end{tabular} & \begin{tabular}{@{}l@{}}Parking\\fee (\$)\end{tabular}& \begin{tabular}{@{}l@{}l@{}}PNR\\parking \\ fee (\$)\end{tabular} \\
        \hline
        1 & 20 & 21 & 10 & 3 \\
        2 & 18 & 22 & 10 & 3 \\
        3 & 22 & 18 & 10 & 3 \\
        4 & 25 & 15 & 10 & 3 \\
        6 (parking) & - & - & - & 3 \\
        \hline
    \end{tabular}
\end{table}

\bibliographystyle{apalike} 
\bibliography{reference}
\end{document}